\documentclass[twocolumn]{article}
\usepackage{graphicx} 
\usepackage{hyperref}

\usepackage{amsmath}
\usepackage[capitalise]{cleveref}
\usepackage{float}
\usepackage{booktabs}
\usepackage[table,xcdraw]{xcolor}
\usepackage{caption}
\newcommand{\method}{DepViT-CAD}
\newcommand{\mavit}{MAViT}
\newcommand{\DGraph}{Diagnosis Distributor Graph}
\newcommand{\dgraph}{D-Graph}

\title{\method{}: Deployable Vision Transformer-Based Cancer Diagnosis in Histopathology}
\author{
Ashkan Shakarami\textsuperscript{1}, 
Lorenzo Nicolè\textsuperscript{2},\\ 
Rocco Cappellesso\textsuperscript{3}, 
Angelo Paolo Dei Tos\textsuperscript{4,5}, 
Stefano Ghidoni\textsuperscript{1}
}

\date{
\small \textsuperscript{1}University of Padova, Italy\\
\small \textsuperscript{2}Unit of Surgical Pathology and Cytopathology, Ospedale dell'Angelo, Italy\\
\small \textsuperscript{3}Pathological Anatomy Unit, Padova University-Hospital, Italy\\
\small \textsuperscript{4}Department of Medicine, University of Padova, Italy\\
\small \textsuperscript{5}Department of Integrated Diagnostics, Azienda Ospedale-Università Padova, Italy\\[1ex]
\small \texttt{
ashkan.shakarami@phd.unipd.it, 
lorenzo.nicole@phd.unipd.it, 
rocco.cappellesso@aopd.veneto.it, 
angelo.deitos@unipd.it, 
stefano.ghidoni@unipd.it
}
}

\begin{document}

\maketitle

\section*{Abstract}
Accurate and timely cancer diagnosis from histopathological slides is vital for effective clinical decision-making. This paper introduces \textbf{\method{}}, a deployable AI system for multi-class cancer diagnosis in histopathology. At its core is \textbf{\mavit{}}, a novel Multi-Attention Vision Transformer designed to capture fine-grained morphological patterns across diverse tumor types. \mavit{} was trained on expert-annotated patches from 1008 whole-slide images, covering 11 diagnostic categories, including 10 major cancers and non-tumor tissue. \method{} was validated on two independent cohorts: 275 WSIs from The Cancer Genome Atlas and 50 routine clinical cases from pathology labs, achieving diagnostic sensitivities of 94.11\% and 92\%, respectively. By combining state-of-the-art transformer architecture with large-scale real-world validation, \method{} offers a robust and scalable approach for AI-assisted cancer diagnostics. To support transparency and reproducibility, \textbf{software and code} will be made publicly available at \href{https://github.com/AshkanShakarami/DepViT-CAD/tree/main}{GitHub}.

\textbf{Keywords:} Cancer Diagnosis, Histopathology, Vision Transformers, Clinical AI Deployment.

\section{Introduction}
\label{sec:HistoCancer_Introduction}
Histological examination of tissue remains the gold standard in modern pathology, offering critical insights to confirm malignancy and guide personalized therapeutic strategies~\cite{Rosai2007, Pisapia2022}. However, the increasing global incidence of cancer has outpaced the availability of qualified pathologists, leading to diagnostic delays and growing clinical workloads~\cite{Metter2019, Markl2021}. This ever-increasing diagnostic gap underscores the urgent need for scalable solutions that can assist pathologists while maintaining diagnostic accuracy and consistency.

The digitization of glass slides into Whole-Slide Images (WSIs) has catalyzed a paradigm shift in pathology. Digital pathology facilitates remote diagnostics and telepathology, while simultaneously enabling the application of advanced computational analysis to histological data~\cite{Betmouni2021, Hanna2022}. In particular, Deep Learning (DL) has emerged as a transformative technology in this domain, demonstrating strong performance in tasks such as tissue classification, tumor detection, and feature segmentation within H\&E-stained slides~\cite{Maier2019, Echle2021, Song2023}. These capabilities have led to the regulatory approval of several AI-based tools for clinical use, including systems for detecting prostate cancer, evaluating Ki-67 and hormonal receptor status in breast cancer, and quantifying PD-L1 expression in lung cancer~\cite{Baxi2022, VanDiest2024}.

Despite this progress, a considerable range of research efforts remain narrowly focused on model development and benchmarking, with limited emphasis on building clinically deployable Computer-Aided Diagnosis (CAD) systems. Such systems serve as critical interfaces between AI algorithms and clinical workflows, enabling real-world translation of computational tools into decision support systems. Addressing this gap, we present \textbf{\method{}}, an end-to-end CAD framework that leverages vision transformer architectures for robust, multi-class cancer diagnosis in histopathology. \method{} is designed to classify 11 histopathological categories, comprising \textbf{10 high-burden tumor types} and \textbf{non-tumor tissue}, across various organ systems. The included malignancies span \textbf{Breast Carcinoma}, \textbf{Non-Small Cell Lung Cancer} (\textit{distinguishing between adenocarcinoma and squamous cell carcinoma}), \textbf{Skin Melanoma}, \textbf{Colon Adenocarcinoma}, \textbf{Gastric Carcinoma}, \textbf{Hepatocarcinoma}, and primary brain tumors including \textbf{Glioblastoma}, \textbf{Astrocytoma}, and \textbf{Oligodendroglioma}.

\textbf{The core challenge addressed by \method{}} lies not in broad organ-level classification, but in the fine-grained recognition of histologically similar tumor subtypes across and within entities that pose substantial diagnostic difficulty in real-world pathology. Rather than distinguishing a brain tumor from a skin tumor---a task that is often straightforward---\method{} targets subtle intra-organ subtype differentiations such as glioblastoma versus astrocytoma, or lung adenocarcinoma versus squamous cell carcinoma, which are known to exhibit significant morphological overlap under H\&E staining. These distinctions frequently challenge even experienced pathologists and require extensive immunohistochemical or molecular analysis. Moreover, the inclusion of non-tumor tissue and early-stage invasive cancers reflects practical clinical scenarios where detecting malignancy within morphologically ambiguous regions is both critical and non-trivial. By addressing these nuanced classification tasks across multiple high-burden cancer types, \method{} contributes meaningfully to the diagnostic workflow and supports timely, subtype-aware decision-making that aligns with current clinical needs.

\textbf{The core contributions of this study} are multifaceted. First, we introduce \method{}, an integrated pipeline for histopathological cancer diagnosis (\autoref{sec:HistoCancer_Methods}). At its core lies \textbf{\mavit{}}, a multi-attention vision transformer classifier specifically designed for high-resolution histopathology image classification (\autoref{sec:HistoCancer_ImageClassification_MAViT}). To further enhance representational capacity, we incorporate a \textbf{Vision Transformer Module (VTM)} (\autoref{sec:HistoCancer_VTM}) and a \textbf{Dual Fusion Strategy (DFS)} that enables effective multi-scale feature integration (\autoref{sec:HistoCancer_DFS}). To facilitate interpretability and clinician trust, we introduce the \textbf{\DGraph{} (\dgraph{})} for intuitive visualization of model predictions (\autoref{fig:2_HistoCancer_D_Graph}, \autoref{sec:HistoCancer_D-Graph}).

\section{Literature Review}
\label{sec:HistoCancer_Literature}

The integration of DL into DP has substantially improved the accuracy, automation, and scalability of cancer diagnosis using WSI~\cite{Shakarami2025thesis}. While traditional machine learning relied on handcrafted features and statistical models with limited generalizability, recent advancements—such as convolutional neural networks (CNNs), vision transformers (ViTs), hybrid models, foundation models, and self-supervised learning (SSL)—have significantly enhanced diagnostic performance in computational pathology. This section reviews key developments in DL-based histopathology, with emphasis on state-of-the-art and emerging methodologies.

CNNs have served as the backbone of DL in pathology, effectively capturing hierarchical spatial features from high-resolution histological data. Architectures like PathCNN and EfficientNet have shown strong performance in tumor detection and subtype classification by leveraging deep features and optimized scaling~\cite{oh2021pathcnn, Tan2019, Shakarami2023, shakarami2024histo, Shakarami2020a,Shakarami2021b,shakarami2021yolov3}. More recent models such as CLAM integrated attention mechanisms into CNNs for weakly supervised WSI classification~\cite{Lu2021}. However, the limited receptive fields of CNNs restrict their ability to capture long-range dependencies, prompting the rise of transformer-based alternatives.

Transformers, through self-attention mechanisms, address this limitation by modeling global contextual relationships. ViT-B16 demonstrated the potential of transformers in WSI classification, albeit requiring large labeled datasets~\cite{Dosovitskiy2020}. DeiT improved data efficiency via distillation~\cite{Touvron2021}, while Swin Transformer employed shifted windows for multi-scale learning~\cite{liu2021swin}. TransMIL and HIPT extended transformer-based learning to multiple-instance and multi-resolution settings, further enhancing classification performance~\cite{Shao2021TransMIL, chen2022HIPT}.

Hybrid models combining CNNs and transformers offer complementary strengths. For instance, TCNN fused convolutional encoders with transformer heads for robust classification of tissue and artifacts~\cite{Shakarami2023}, while other hybrid networks employed multi-scale fusion to improve generalization across tumor types~\cite{Li2023}.

Foundation models have further shifted the field by offering pre-trained architectures adaptable to diverse pathology tasks. The UNI model, trained on over 100 million image tiles, exemplifies the scalability and generalizability of self-supervised approaches~\cite{Chen2023}. Models such as CONCH and mSTAR have demonstrated the value of multimodal learning by combining histological and genomic data to improve classification and biomarker discovery~\cite{CONCH, mSTAR}.

SSL techniques have gained momentum as solutions to annotation scarcity. Contrastive learning methods like SimCLR and MoCo have enabled models to learn robust features from unlabeled WSIs~\cite{SimCLR, MoCo}. Meanwhile, approaches like Pathomic Fusion combine histology and omics-level data, supporting more comprehensive diagnostic frameworks~\cite{chen2020pathomic}.

Despite these developments, key challenges remain. Transformer models still demand extensive labeled data and impose high computational costs, limiting their clinical scalability. Integrating heterogeneous data modalities—histological, genomic, and clinical—remains technically complex. Additionally, interpretability continues to be a major hurdle for clinical adoption, as transparency and trust are essential for AI acceptance in pathology. Addressing these limitations will require lightweight, generalizable, and clinically interpretable models to fully realize the potential of AI-assisted cancer diagnostics.

\section{\method{}}
\label{sec:HistoCancer_Methods}

\begin{figure}[h!]
    \centering
    \includegraphics[angle=0, width=0.5\textwidth]{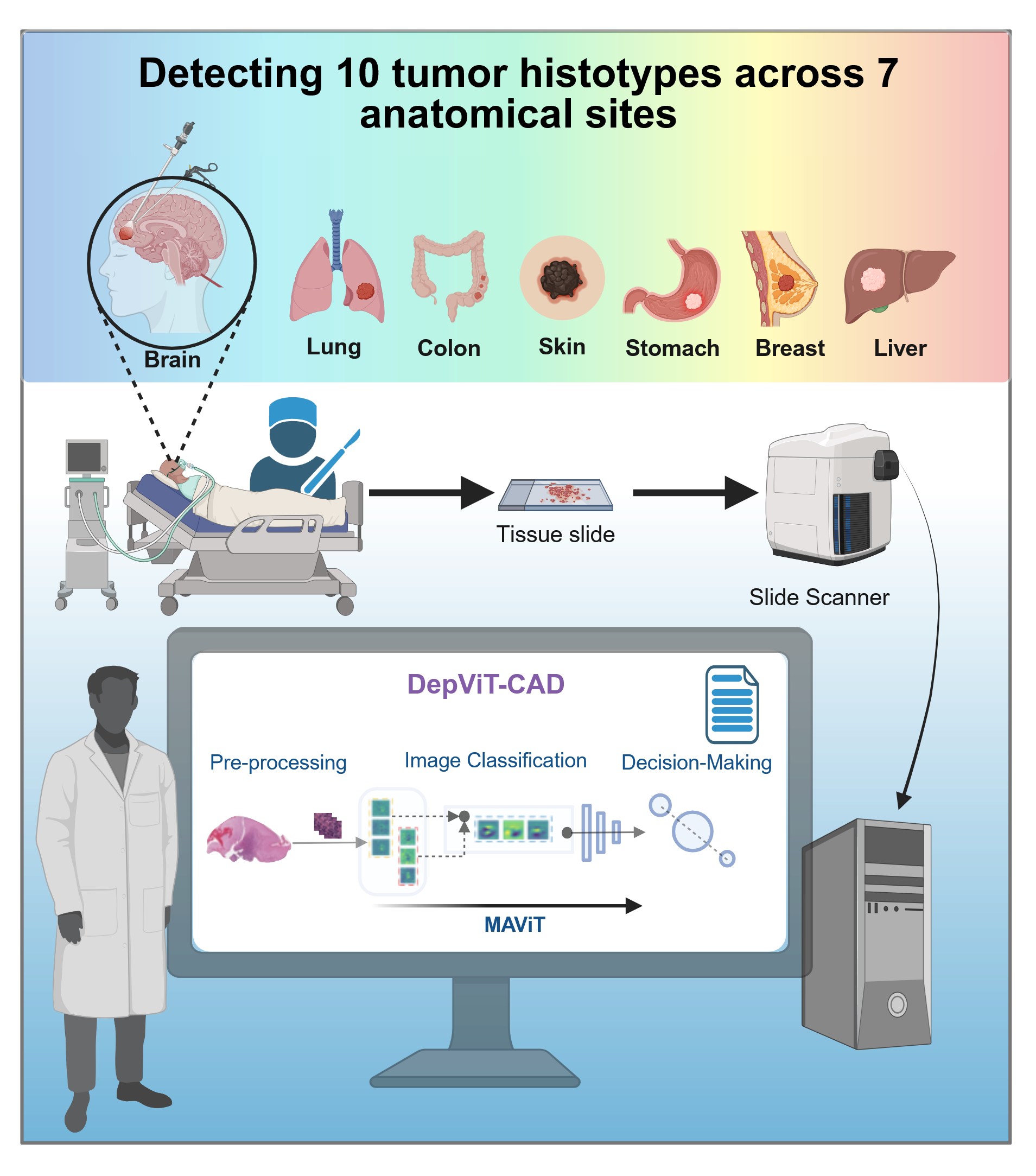}
    \caption{Cancer Diagnostic Pipeline via \method{}}
    \label{fig:HistoCancer-CAD}
\end{figure}

The diagnostic pipeline, summarized in ~\autoref{fig:HistoCancer-CAD}, comprises tissue preparation, digital scanning, and computational inference using the MAViT architecture—a vision transformer-based model optimized for high-resolution histology images. Each phase is described in the following sections.

\subsection{Pre-processing}
\label{sec:HistoCancer_Pre-processing}

To ensure flexibility in both development and deployment, the pre-processing pipeline was divided into two phases: model training and clinical application.

\subsubsection{Model Training Phase}
\label{sec:HistoCancer_TrainingPreprocessing}

In the training phase, WSIs are processed using the SlideTiler tool~\cite{Barcellona2024}. A certified pathologist manually annotates tumor and non-tumor regions at 40$\times$ magnification. These annotated ROIs, denoted as \( R_i \), are subdivided into non-overlapping RGB patches of size $512 \times 512 \times 3$ pixels. The tiling process is defined by:

\[
I_i = \bigcup_{j=1}^{N} T_{ij} \tag{3.1}
\label{eq:HistoCancer_Preprocessing}
\]

where \( I_i \) is the input region and \( T_{ij} \) denotes the \( j^{th} \) tile. For instance, a region of size $17920 \times 12160$ yields approximately \( \lceil17920/512\rceil \times \lceil12160/512\rceil \) tiles. To maintain computational tractability, a representative subset of 100 tiles per patient was randomly sampled. This choice strikes a balance between computational efficiency and morphological diversity and minimizes the model’s exposure to global tissue architecture, encouraging tumor-specific feature learning over organ-specific patterns.

\subsubsection{Clinical Deployment Phase}
\label{sec:HistoCancer_ClinicalPreprocessing}

For clinical use, the model supports two operational modes:

\begin{itemize}
    \item \textbf{Full WSI Analysis:} The entire WSI is tiled and processed automatically.  
    \item \textbf{ROI-Based Analysis:} A user-specified region is selected for localized diagnostic focus.
\end{itemize}

This dual-mode functionality accommodates variable diagnostic workflows and ensures that the system adapts to both broad and focused diagnostic requirements.

\subsubsection{Discussion on Pre-Processing}

In deployment, HistoCancer-CAD operates independently of manual annotations. It processes entire WSIs or user-defined ROIs, generalizing to unseen tissue regions. Training on both tumor and non-tumor patches enhances the model’s robustness to variable tissue morphology.

In clinical settings, automated inference has demonstrated substantial diagnostic utility. For example, during the evaluation of a glioma case where histological review could not resolve subtype ambiguity, HistoCancer-CAD correctly predicted Astrocytoma within minutes. The final molecular diagnosis later confirmed the model’s output. While not a substitute for molecular diagnostics, HistoCancer-CAD offers preliminary guidance to accelerate clinical decision-making.

To mitigate variability in pathologist annotations, the training dataset was curated by multiple board-certified pathologists. This ensured consistency and minimized inter-observer bias. During deployment, the model bypasses such manual steps altogether, improving reproducibility and operational independence. Experimental evaluations on the CliniR-ds dataset confirm that the model delivers consistent performance across all diagnostic categories, regardless of input mode, further validating its generalizability.

\subsection{Image Classification with \mavit{}}
\label{sec:HistoCancer_ImageClassification_MAViT}
\begin{figure}[h!]
    \centering
    \includegraphics[width=0.5\textwidth]{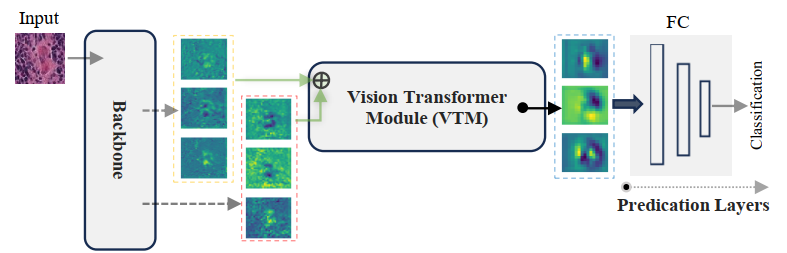}
    \caption{\mavit{}’s overview (Linked to Figure \ref{fig:11_HistoCancer_MAViT_Architecture})}
    \label{fig:4_HistoCancer_MAViT_Overview}
\end{figure}

\begin{figure*}[h!]
\centering
\includegraphics[width=0.7\textwidth]{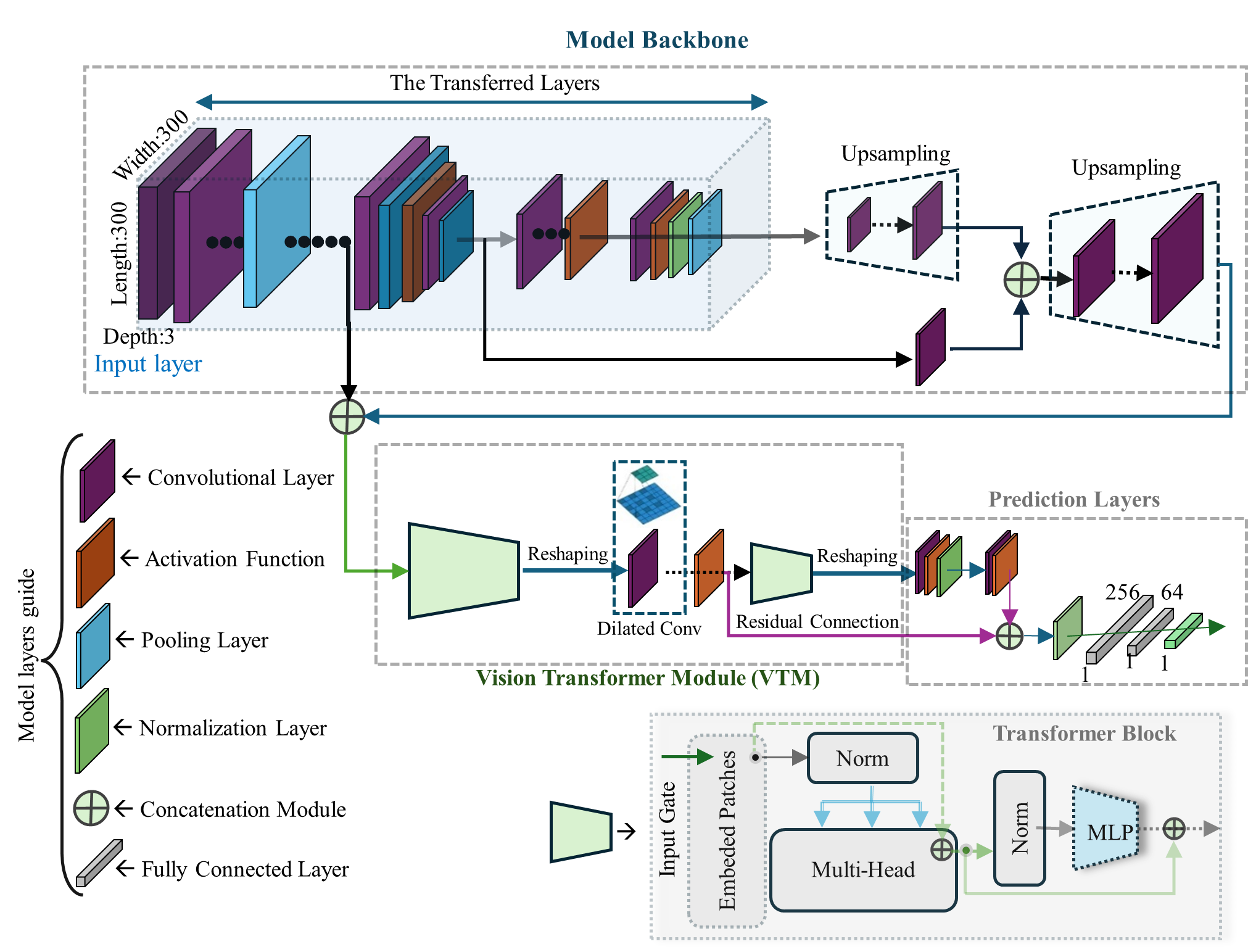}
\caption{\mavit{} Architecture; integrating Multi-Attention Transformers, Multi-Layer Feature Fusion, and Convolutional Layers for Histopathological Image Classification}
\label{fig:11_HistoCancer_MAViT_Architecture}
\end{figure*}

\mavit{} introduces an advanced approach for histopathological image classification by integrating convolutional and Transformer-based architectures. \autoref{fig:4_HistoCancer_MAViT_Overview} outlines its core components, with a detailed architectural view shown in \autoref{fig:11_HistoCancer_MAViT_Architecture}. The backbone combines pre-trained layers from EfficientNet-B3 with additional convolutional blocks to extract rich, hierarchical features from input patches. VTM incorporates self-attention layers to model contextual relationships across spatial regions. Finally, the prediction head consolidates the learned features and performs multi-class classification through fully connected layers and softmax activation.

\autoref{fig:11_HistoCancer_MAViT_Architecture} presents a detailed illustration of the \mavit{} architecture. The upper section highlights the model's backbone, built upon transferred layers from EfficientNet-B3, enhanced with upsampling modules and skip connections to enable multi-level feature fusion. Among the EfficientNet variants, B3 was chosen for its balance between classification accuracy and computational efficiency. The Multi-Level Feature Fusion module integrates representations from various depths of the network, enabling the model to capture both fine-grained local patterns and high-level semantic features, thereby enhancing its diagnostic performance.

\subsubsection{Backbone}
\label{sec:HistoCancer_Backbone}

The backbone of \mavit{} is based on EfficientNet-B3, a CNN architecture that applies compound scaling to balance depth, width, and input resolution~\cite{Tan2019}. This strategy enables efficient performance across varied tasks. In this framework, EfficientNet-B3 is used without modification to extract hierarchical features from histopathological image patches. The resulting feature map has a spatial resolution of $32 \times 32$ with 128 channels, which is subsequently processed by VTM to capture contextual relationships via self-attention.

\subsubsection{VTM}
\label{sec:HistoCancer_VTM}

VTM enhances \mavit{}'s capacity to model both local and long-range dependencies within histopathological feature maps. While convolutional layers effectively capture local textures and spatial structures, they are inherently limited in representing relationships across distant regions. To overcome this, VTM incorporates self-attention mechanisms capable of learning global context.

\begin{figure}[h!]
    \centering
    \includegraphics[width=0.49\textwidth]{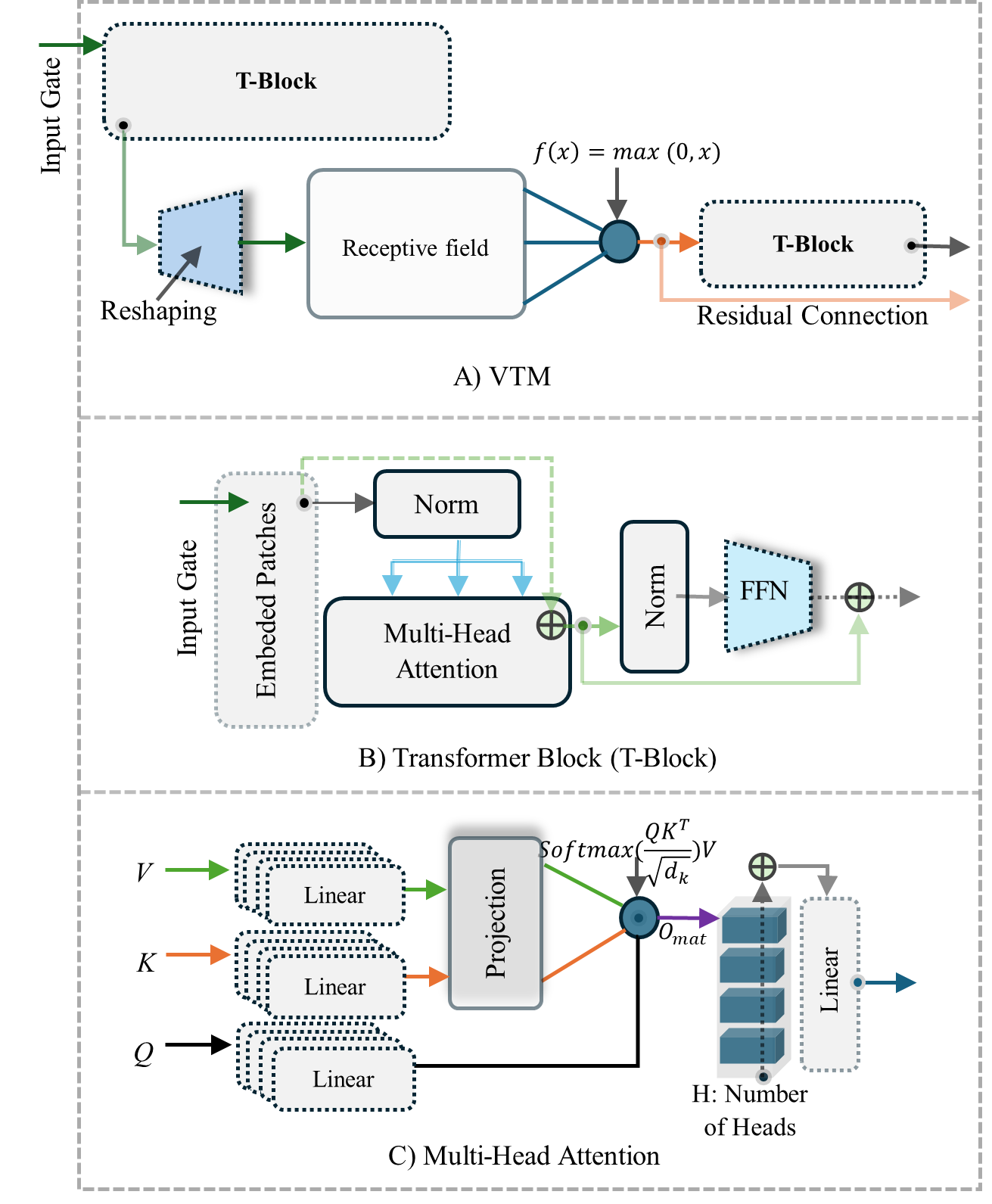}
    \caption{Architecture of the VTM \cite{Dosovitskiy2020}}
    \label{fig:6_HistoCancer_VTM}
\end{figure}

As depicted in \autoref{fig:6_HistoCancer_VTM}, VTM comprises three key components: (A) a residual connection path to facilitate gradient flow, (B) a Transformer Block (T-Block) with attention layers for capturing non-local interactions, and (C) a dilated convolutional layer that expands the receptive field without significantly increasing computational cost.

The T-Block, illustrated in \autoref{fig:6_HistoCancer_VTM} (B), adopts the architectural principles of standard Transformer designs. It integrates multi-head attention to capture spatially distributed relationships across the feature map, enabling the model to focus on contextually relevant regions. A feed-forward network composed of fully connected layers with non-linear activations transforms the intermediate representations. Normalization layers are employed to stabilize gradient flow and ensure consistent feature scaling during training. Residual connections are incorporated to preserve input information across layers, facilitating effective backpropagation and improving convergence stability.

To address the high computational cost typically associated with self-attention, the VTM incorporates a Multi-Head Linear Attention mechanism inspired by the Linformer architecture~\cite{Linformer2020}. This technique reduces the complexity of the attention operation by projecting the query and key matrices into lower-dimensional representations, effectively transforming the computational burden from quadratic to linear with respect to input size. The linear attention formulation is expressed as \( O_{\text{mat}} = \text{Softmax} \left( \frac{QK^T}{\sqrt{d_k}} \right) V \), where \( Q \), \( K \), and \( V \) denote the query, key, and value matrices, and \( d_k \) is the dimensionality of the key vectors.

By combining this attention mechanism with convolutional processing, the VTM allows the network to leverage both local features and broader contextual information. The impact of this hybrid design is quantitatively assessed in the ablation study presented in \autoref{sec:Ablation_Study}.

\subsubsection{Prediction Layers}
\label{sec:HistoCancer_Prediction_layers}
The prediction layers in the \mavit{} architecture convert the contextualized features from the VTM into final classification outputs. The process begins with a convolutional layer followed by a ReLU activation to enhance salient spatial patterns, with normalization applied to stabilize training. A second convolution and normalization layer further refine the representation. To preserve the original semantic context, a residual connection merges the initial VTM output with the refined features. The resulting feature map is then passed through fully connected layers, concluding with a softmax activation that yields the final probability distribution across the target classes.

\subsubsection{DFS}
\label{sec:HistoCancer_DFS}
\begin{figure}[h!]
    \centering
    \includegraphics[width=0.3\textwidth]{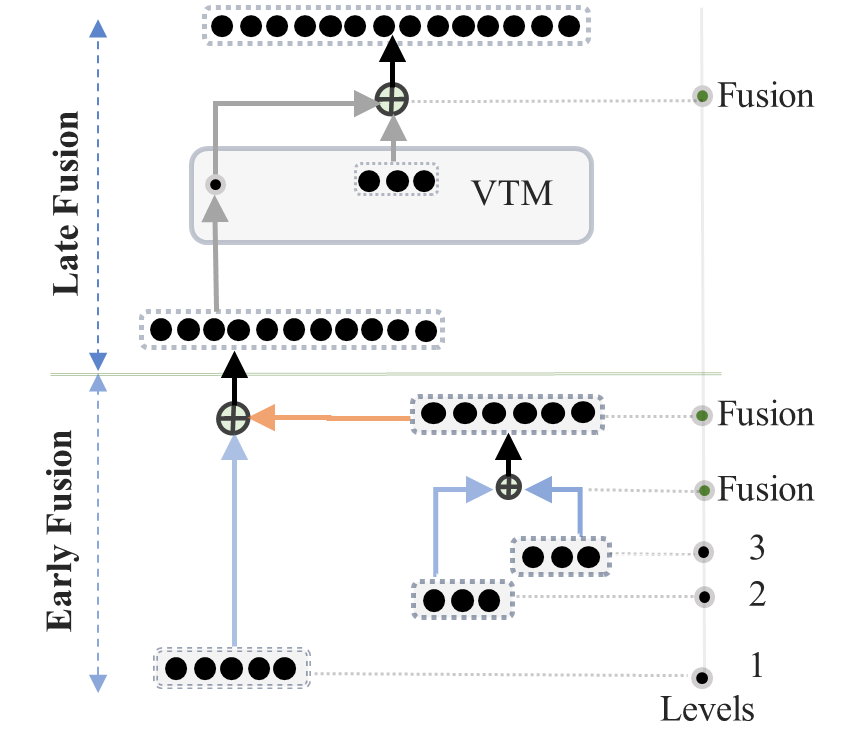}
    \caption{DFS integrates early and late fusion for hierarchical feature processing.  Early fusion (lower part) combines multi-level features at Levels 1–3 within \mavit{} (\autoref{fig:4_HistoCancer_MAViT_Overview} and \autoref{fig:11_HistoCancer_MAViT_Architecture}), while late fusion (upper part) aggregates final-stage features before and after passing them to the VTM \autoref{fig:6_HistoCancer_VTM}. The $\oplus$ symbol denotes the fusion operation.}
    \label{fig:8_HistoCancer_DFS}
\end{figure}

To strengthen feature representation, \mavit{} incorporates a DFS that merges multi-scale features extracted from different depths of the EfficientNet-B3 backbone. This hierarchical fusion enables the model to leverage both fine-grained local details and abstract semantic information in the classification process. As illustrated in \autoref{fig:8_HistoCancer_DFS}, the DFS operates in two stages. The Early Fusion stage aggregates features from multiple backbone layers (\autoref{fig:12_HistoCancer_MAViT_CAM}), while the Late Fusion stage combines these enriched features with contextual representations from the VTM, ensuring comprehensive information flow across the network.

\paragraph{Early Fusion:} 
In the Early Fusion stage, hierarchical features $F_{\text{backbone}}^{(l)}$ are extracted from three levels of the EfficientNet-B3 backbone, each representing different depths of feature abstraction:
\begin{itemize}
    \item \textbf{Shallow Features (Low-Level Representation)}: Capturing fine-grained details with a feature map size of $32\times32\times64$.
    \item \textbf{Intermediate Features (Mid-Level Representation)}: Extracting structural and contextual patterns with a feature map size of $18\times18\times64$.
    \item \textbf{Deep Features (High-Level Representation)}: Encoding abstract and semantic information with a feature map size of $10\times10\times64$.
\end{itemize}

To ensure consistent spatial dimensions prior to merging, the deep features are first passed through a \(2 \times 2\) convolutional layer, followed by bilinear upsampling to double their spatial resolution, aligning them with the intermediate features. The intermediate features undergo a \(1 \times 1\) convolution for refinement and are then resized to \(18 \times 18\) to match the processed deep features. Finally, the shallow features are upsampled and concatenated with the aligned intermediate and deep features, forming the complete Early Fusion representation.

This structured feature merging process is formulated as \autoref{eq:HistoCancer_Early_Fusion}, where $\bigoplus$ denotes the fusion operation combining different-scale feature maps after spatial alignment.

\begin{equation}
F_{\text{early}} = \bigoplus_{l=3}^{5} F_{\text{backbone}}^{(l)} 
\tag{3.4}
\label{eq:HistoCancer_Early_Fusion}
\end{equation}

\paragraph{Late Fusion:}
After the Early Fusion stage, the combined feature set \( F_{\text{early}} \) undergoes further refinement through Late Fusion. This phase integrates the multi-scale features derived from the CNN backbone with the global contextual representations learned by VTM, denoted as \( F_{\text{VTM}} \). As defined in \autoref{eq:HistoCancer_Late_Fusion}, this fusion mechanism allows the model to jointly leverage local spatial hierarchies and long-range dependencies, enhancing the overall representational capacity for accurate classification.

\begin{equation}
F_{\text{Late}} = F_{\text{early}} \bigoplus F_{\text{VTM}} 
\tag{3.5}
\label{eq:HistoCancer_Late_Fusion}
\end{equation}

\paragraph{Comparison with ResNet and DenseNet:}
ResNet and DenseNet are prominent convolutional architectures known for their use of residual and densely connected layers, which aid in feature propagation and gradient flow. However, neither architecture explicitly addresses the spatial alignment of features across different resolutions prior to fusion. ResNet aggregates features by adding outputs from earlier layers to later ones, while DenseNet promotes feature reuse by concatenating features from all preceding layers. In contrast, \mavit{} employs a DFS that incorporates bilinear up-sampling and structured spatial resizing before merging multi-level features. This alignment ensures consistency in feature dimensions and provides a unified representation to the Transformer module, improving the model’s ability to integrate both low-level and high-level contextual information.

\textbf{Characteristics of DFS:}
DFS combines multi-scale feature representations extracted from various depths of the network, ensuring the integration of both fine-grained local patterns and high-level semantic information. Before fusion, spatial dimensions are aligned using resizing operations, which prevents inconsistencies that may arise from merging features of differing resolutions. This structured alignment enhances the coherence of the fused representation and supports more effective downstream processing.

\subsection{Decision-Making}
\label{sec:HistoCancer_Decision}
\begin{figure*}[h!]
    \centering
    \includegraphics[width=0.9\textwidth]{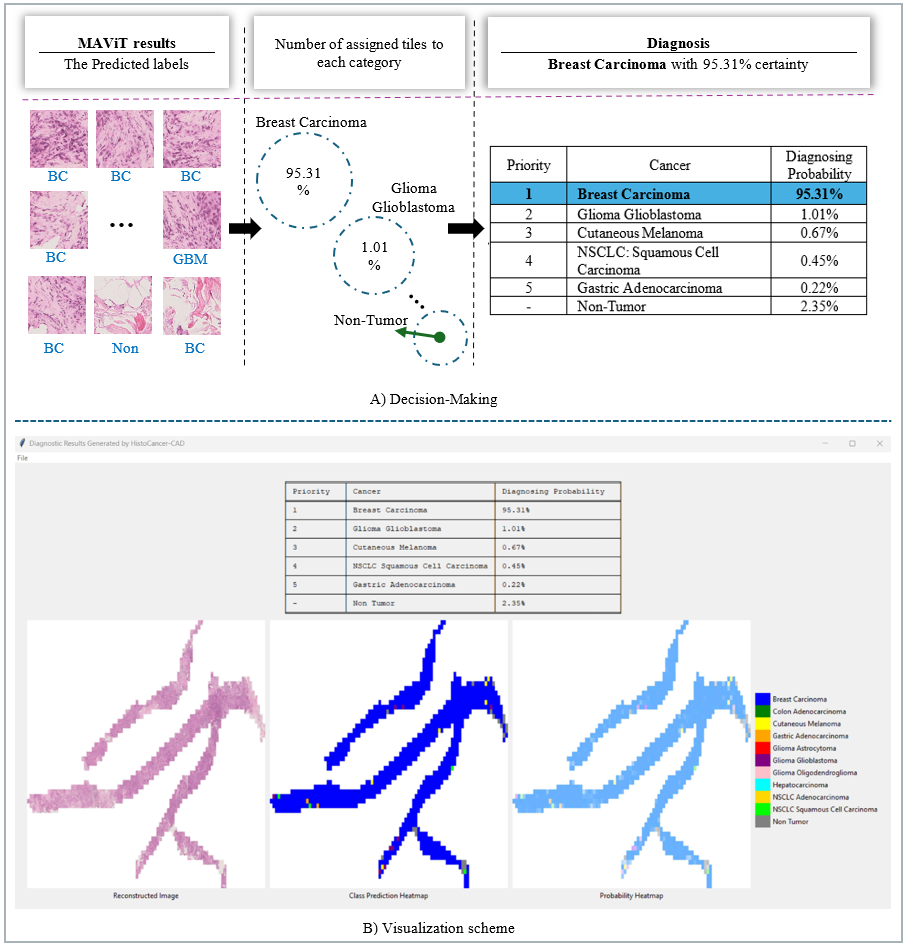}
    \caption{\method{}’s Decision-Making and Visualization scheme. The output includes tumor and non-tumor predictions within the same tissue sample.}
    \label{fig:9_HistoCancer_Decision_Making}
\end{figure*}

The final stage of the \method{} pipeline involves aggregating patch-level predictions to produce a region-level or whole-slide diagnosis. As illustrated in \autoref{fig:9_HistoCancer_Decision_Making}, this aggregation is performed using a majority voting scheme, where individual patch classifications are combined to yield a single diagnostic outcome. This strategy enables robust interpretation of localized predictions and reduces the influence of isolated misclassifications.

The classification task spans 11 diagnostic categories, including ten tumor types and one non-tumor class. The tumor classes encompass three glioma subtypes (Glioblastoma, Astrocytoma, and Oligodendroglioma), two forms of lung cancer (Adenocarcinoma and Squamous Cell Carcinoma), and five additional cancers: Cutaneous Melanoma, Colon Adenocarcinoma, Gastric Adenocarcinoma, Breast Carcinoma, and Hepatocarcinoma. The non-tumor category accounts for healthy or morphologically normal tissue regions, enhancing the system’s ability to distinguish pathological from benign features.

\method{} can be applied to either an entire WSI or a targeted ROI, depending on the clinical context and user preference. This flexibility enables comprehensive or localized diagnostic assessment. Representative visualizations of the system’s outputs are provided in \autoref{fig:9_HistoCancer_Decision_Making}, \autoref{fig:3_HistoCancer_Visual_Output}, and \autoref{fig:Gastric}.

As illustrated in \autoref{fig:9_HistoCancer_Decision_Making}, the table in Sub-figure (B) reveals that the model assigns both tumor and non-tumor labels to distinct regions within the same WSI, demonstrating its ability to differentiate heterogeneous tissue types across a single slide. The brown-highlighted areas, corresponding to non-tumor predictions as per the GUI legend, indicate that \method{} is capable of identifying morphologically normal tissue even in slides dominated by malignant regions, such as this Breast Carcinoma specimen.

Further evidence of this distinction is provided in \autoref{fig:3_HistoCancer_Visual_Output} and \autoref{fig:Gastric}, where the model accurately localizes tumor regions while sparing adjacent non-tumor tissue. In the case of Cutaneous Melanoma, for example, the model successfully isolates neoplastic areas without conflating them with surrounding healthy skin structures. This behavior suggests that \method{} does not rely on organ-specific patterns alone, but rather learns to identify tumor-specific features within a given histological context.

The ability to segment ROIs containing both tumor and non-tumor areas further supports this claim. If the model were biased toward organ classification, it would likely assign a homogeneous label based on tissue origin. Instead, the differential labeling observed across these examples underscores the system’s sensitivity to pathological features, validating its tumor-centric rather than organ-centric learning paradigm.

\subsubsection{Aggregating Predictions}
\label{sec:HistoCancer_Aggregating Predictions}
Let \( T = \{t_1, t_2, \dots, t_n\} \) represent the set of \( n \) patches extracted from a WSI. For each patch \( t_i \), the model generates a predicted class label \( \hat{y}_i \in Y \), where \( Y \) is the set of possible cancer labels (e.g., Breast Carcinoma, Lung Adenocarcinoma, etc.). The decision-making process then calculates the proportion \( p_y \) of patches classified under each label \( y \in Y \), as shown in Equation \ref{eq:tile_proportion}.

\begin{equation}
    p_y = \frac{1}{n} \sum_{i=1}^{n} \delta(\hat{y}_i, y)
    \label{eq:tile_proportion}
\end{equation}

Where \( \delta(\hat{y}_i, y) \) is the Kronecker delta function, defined in \autoref{eq:kronecker_delta}:

\begin{equation}
    \delta(\hat{y}_i, y) = 
    \begin{cases}
        1, & \text{if } \hat{y}_i = y \\
        0, & \text{if } \hat{y}_i \neq y
    \end{cases}
    \label{eq:kronecker_delta}
\end{equation}

The final diagnosis \( y^* \) is determined by majority voting, selecting the label with the highest proportion \( p_y \), as expressed in \autoref{eq:final_diagnosis}.

\begin{equation}
    y^* = \arg\max_{y \in Y} p_y
    \label{eq:final_diagnosis}
\end{equation}

For example, if 95.31\% of tiles are classified as Breast Carcinoma (BC), the final diagnosis \( y^* \) is Breast Carcinoma with a certainty of 95.31\%.

The current approach employs a hard-decision voting scheme based on the Kronecker delta function to aggregate patch-level predictions. While this method has consistently yielded strong performance across experiments—outperforming or matching several baseline and state-of-the-art models—it carries inherent limitations in borderline classification scenarios. Specifically, when two or more classes exhibit similar confidence scores, the discrete nature of the delta function enforces a binary outcome, which may reduce the robustness of whole-slide predictions in ambiguous regions.

An alternative strategy would involve soft-label aggregation, wherein softmax probabilities are accumulated across all patches to form a probabilistic estimate of the WSI-level label~\cite{huang2024learnable, couture2018multiple}. This approach could enable more calibrated decision-making and improve resilience to uncertainty in heterogeneous slides.

Given the demonstrated accuracy and clinical alignment of the current model—validated on both the TCGA and CliniR datasets—revisions to the aggregation scheme are reserved for future work. Future research may explore the integration of soft-label aggregation to enhance diagnostic stability, particularly in cases characterized by histological ambiguity or class overlap. Nonetheless, the present methodology remains effective and clinically viable within the scope of this study.

\subsubsection{Visual Analysis}
\label{sec:HistoCancer_Visual Analysis}
After applying the majority voting scheme, \method{} generates visual outputs using a GUI, as shown in \autoref{fig:9_HistoCancer_Decision_Making} (part b). The interface includes a table listing non-zero predicted probabilities for each cancer class, along with three supporting visualizations:

\begin{enumerate}
    \item \textbf{Reconstructed Input Image}: A composite view of the original WSI based on the tiled inputs used for prediction.
    \item \textbf{Class Prediction Labelmap}: A visualization of predicted class labels, indicating the spatial location of each assigned class within the WSI.
    \item \textbf{Probability Heatmap}: A representation of the prediction confidence levels, with pixel intensities corresponding to class probability scores.
\end{enumerate}

These visual outputs provide a spatial summary of the model’s patch-wise predictions. They are intended to support further inspection of classification results by users, including pathologists or researchers, particularly in cases where regional variation in prediction confidence may be relevant.

\subsubsection{Model Predictions Measurement by \dgraph{}}
\label{sec:HistoCancer_D-Graph}
The \dgraph{}, shown in \autoref{fig:2_HistoCancer_D_Graph}, is a visualization designed to organize model predictions based on classification correctness and sample modality (surgical vs. biopsy). While traditional confusion matrices provide a numerical summary of prediction outcomes, they do not reflect whether samples originate from surgical or biopsy procedures. To complement this limitation, the \dgraph{} was introduced to display both prediction results and sample modalities in a unified visual format.

Let \( S = \{s_1, s_2, \dots, s_N\} \) denote the set of \( N \) clinical samples, where each sample \( s_i \) has a modality label \( m_i \in \{\text{Surgical}, \text{Biopsy}\} \), and a corresponding set of extracted image tiles \( I(s_i) \). Predictions are generated via a classification function \( f: S \rightarrow \{\text{Positive}, \text{Negative}\} \), where \( \hat{y}_i = f(s_i) \) denotes the predicted label and \( y_i \) is the ground truth.

\paragraph{Quadrant-Based Structure}
The \dgraph{} uses a quadrant layout to distinguish prediction correctness and modality type:
\begin{itemize}
\item Top-left: Correct predictions on biopsy samples
    \item Top-right: Correct predictions on surgical samples
    \item Bottom-left: Incorrect predictions on biopsy samples
    \item Bottom-right: Incorrect predictions on surgical samples
\end{itemize}

Each clinical sample is represented as a circle, with its size scaled proportionally to the number of extracted tiles \( I(s_i) \).

\paragraph{Sample Positioning in the Graph}
The horizontal coordinate \( x_i \) is determined by the sample modality:

\begin{equation}
x_i =
\begin{cases}
 +|x_{\text{max}}| \times \left(\frac{i}{N}\right), & \text{if } m_i = \text{Surgical} \\
 -|x_{\text{max}}| \times \left(\frac{i}{N}\right), & \text{if } m_i = \text{Biopsy}
\end{cases}
\tag{3.9}
\label{eq:HistoCancer_Sample_Position}
\end{equation}

where \( x_{\text{max}} \) is the maximum horizontal value. The vertical coordinate is determined by prediction correctness and the number of tiles per sample:

\begin{equation}
\hat{y}_i =
\begin{cases}
 +|y_{\text{max}}| \times \left(\frac{I(s_i)}{I_{\text{max}}}\right), & \text{if } \hat{y}_i = y_i \\
 -|y_{\text{max}}| \times \left(\frac{I(s_i)}{I_{\text{max}}}\right), & \text{if } \hat{y}_i \neq y_i
\end{cases}
\tag{3.10}
\label{eq:HistoCancer_Vertical_Position}
\end{equation}

Here, \( y_{\text{max}} \) defines the vertical axis range, and \( I_{\text{max}} \) is the maximum number of image tiles observed across all samples.

\paragraph{\dgraph{} Interpretation}  
\dgraph{} is a complementary visualization tool alongside conventional evaluation metrics such as the confusion matrix. Beyond indicating classification accuracy, it facilitates the analysis of diagnostic trends with respect to variables like sample type (e.g., biopsy vs. surgical specimens) and the number of extracted tiles per case. This visualization supports deeper insight into model behavior across different clinical contexts and case complexities.

\section{Results}
\subsection{Dataset and Training Strategy}
\label{sec:HistoCancer_Dataset_Training}
\label{sec:HistoCancer_Results}

\subsubsection{Dataset Acquisition}
\label{sec:HistoCancer_Datasets}
The dataset used in this study comprises WSIs from two primary sources: the publicly available TCGA repository and a real-world clinical dataset, CliniR-ds, curated from the Angelo Hospital Pathology Unit in Mestre (Venice). The TCGA dataset, referred to as TgCancer-ds, includes 1283 WSIs from an equal number of patients representing ten cancer types and non-tumor cases (\autoref{tab:HistoCancer_Tgcancer_ds_info}). The CliniR-ds consists of 50 retrospective, anonymized cases covering the same ten cancer categories, including biopsy and surgical specimens. WSIs from CliniR-ds were scanned at 40× magnification using a Ventana DP200 scanner (Ventana, Roche), and all cases were meticulously evaluated by a board-certified pathologist (\autoref{tab:HistoCancer_CliniRDs_eval}).

As mentioned earlier, the \method{} has been designed to process either a WSI or an ROI based on the pathologist's selection. This flexibility allows for the presence of non-tumor (healthy) tissue within the analyzed samples, which has been considered in our experiments. The visual outputs further support this (\autoref{fig:3_HistoCancer_Visual_Output}, \autoref{fig:Gastric}).

All experimental procedures adhered to the ethical guidelines outlined in the 1964 Helsinki Declaration (and its subsequent revisions), and the development of diagnostic models followed the TRIPOD (Transparent Reporting of a Multivariate Prediction Model for Individual Prognosis or Diagnosis) guidelines \cite{Collins2015}.

\subsubsection{Dataset Partitioning}
\label{sec:HistoCancer_Dataset_Partitioning}
For model development, the TgCancer-ds was partitioned at the patient-level, ensuring that all WSIs from a single patient were exclusively allocated to either the training, validation, or test set. This approach prevents data leakage and ensures robust generalization. Specifically, the training set comprised 869 patients (the 67.8\% of all data used), the validation set included 139 patients (10.8\%), and the test set contained 275 patients (21.4\%) (\autoref{tab:HistoCancer_Tgcancer_ds_info}). The CliniR-ds was used to evaluate the generalization capability of the model further for external validation. Unlike TgCancer-ds, which consists solely of surgical specimens, CliniR-ds includes both biopsy and surgical samples, providing a realistic validation scenario.

\begin{table}[h!]
\centering
\caption{TgCancer-ds information}
\label{tab:HistoCancer_Tgcancer_ds_info}
\scalebox{0.7}{
\begin{tabular}{|l|c|c|c|c|}
\hline
\textbf{Type}                      & \textbf{Training} & \textbf{Val} & \textbf{Test} & \textbf{Total} \\ \hline
Breast Carcinoma                   & 81                & 13                  & 26           & 120                 \\ \hline
Colon Adenocarcinoma               & 90                & 14                  & 26           & 130                 \\ \hline
Cutaneous Melanoma                 & 67                & 11                  & 23           & 101                 \\ \hline
Gastric Adenocarcinoma             & 85                & 14                  & 21           & 120                 \\ \hline
Glioma Astrocytoma                 & 64                & 10                  & 28           & 102                 \\ \hline
Glioma Glioblastoma                & 72                & 12                  & 25           & 109                 \\ \hline
Glioma Oligodendroglioma           & 68                & 11                  & 19           & 98                  \\ \hline
Hepatocarcinoma                    & 90                & 15                  & 29           & 134                 \\ \hline
NSCLC: Adenocarcinoma              & 88                & 14                  & 29           & 131                 \\ \hline
NSCLC: Squamous Cell Carcinoma     & 90                & 15                  & 29           & 134                 \\ \hline
Non-Tumor                          & 74                & 10                  & 20           & 104                 \\ \hline
\textbf{Patients frequency}        & \textbf{869}      & \textbf{139}        & \textbf{275} & \textbf{1283}        \\ \hline
\textbf{Volume of data}            & \multicolumn{4}{c|}{\textbf{246 gigabytes}}                    \\ \hline
\textbf{Total number of tiles}     & \multicolumn{4}{c|}{\textbf{512271}}                           \\ \hline
\end{tabular}
}
\end{table}

\begin{table*}[h!]
\centering
\caption{Evaluation of \method{} on CliniR-ds with Surgical and Biopsy Samples (S). The numbers denote the image count per patient. Blue cells represent correctly diagnosed biopsy cases, while green cells indicate accurately diagnosed surgical cases. Dark and bright orange cells highlight misdiagnosed surgical and biopsy cases, respectively.}
\label{tab:HistoCancer_CliniRDs_eval}
\scalebox{0.8}{
\begin{tabular}{|l|c|c|c|c|c|}
\hline
\textbf{Cancer Type}             & \textbf{S 1} & \textbf{S 2} & \textbf{S 3} & \textbf{S 4} & \textbf{S 5} \\ \hline
Breast Carcinoma         & \cellcolor[HTML]{C9DAF8}266  & \cellcolor[HTML]{C9DAF8}596  & \cellcolor[HTML]{C9DAF8}887  & \cellcolor[HTML]{C9DAF8}895  & \cellcolor[HTML]{C9DAF8}1006 \\ \hline
Colon Adenocarcinoma     & \cellcolor[HTML]{9AFF99}53   & \cellcolor[HTML]{9AFF99}91   & \cellcolor[HTML]{9AFF99}287  & \cellcolor[HTML]{E06666}653  & \cellcolor[HTML]{9AFF99}157  \\ \hline
Cutaneous Melanoma       & \cellcolor[HTML]{9AFF99}1862 & \cellcolor[HTML]{9AFF99}1022 & \cellcolor[HTML]{9AFF99}78   & \cellcolor[HTML]{9AFF99}701  & \cellcolor[HTML]{9AFF99}493  \\ \hline
Gastric Adenocarcinoma                            & \cellcolor[HTML]{F4CCCC}691  & \cellcolor[HTML]{9AFF99}216  & \cellcolor[HTML]{9AFF99}6955 & \cellcolor[HTML]{9AFF99}5505 & \cellcolor[HTML]{9AFF99}2387 \\ \hline
Glioma Astrocytoma                                & \cellcolor[HTML]{E06666}4002 & \cellcolor[HTML]{9AFF99}4141 & \cellcolor[HTML]{9AFF99}465  & \cellcolor[HTML]{9AFF99}4711 & \cellcolor[HTML]{9AFF99}5374 \\ \hline
Glioma Glioblastoma                               & \cellcolor[HTML]{9AFF99}2070 & \cellcolor[HTML]{9AFF99}2096 & \cellcolor[HTML]{9AFF99}3511 & \cellcolor[HTML]{9AFF99}2719 & \cellcolor[HTML]{9AFF99}2825 \\ \hline
Glioma Oligodendroglioma                          & \cellcolor[HTML]{9AFF99}3192 & \cellcolor[HTML]{9AFF99}3316 & \cellcolor[HTML]{9AFF99}2719 & \cellcolor[HTML]{9AFF99}2604 & \cellcolor[HTML]{9AFF99}1782 \\ \hline
Hepatocarcinoma                                   & \cellcolor[HTML]{9AFF99}676  & \cellcolor[HTML]{9AFF99}922  & \cellcolor[HTML]{9AFF99}1357 & \cellcolor[HTML]{9AFF99}650  & \cellcolor[HTML]{9AFF99}2532 \\ \hline
NSCLC: Adenocarcinoma                             & \cellcolor[HTML]{E06666}2415 & \cellcolor[HTML]{C9DAF8}410  & \cellcolor[HTML]{C9DAF8}863  & \cellcolor[HTML]{9AFF99}2163 & \cellcolor[HTML]{9AFF99}5010 \\ \hline
NSCLC: Squamous Cell Carcinoma                    & \cellcolor[HTML]{C9DAF8}419  & \cellcolor[HTML]{C9DAF8}420  & \cellcolor[HTML]{C9DAF8}53   & \cellcolor[HTML]{C9DAF8}261  & \cellcolor[HTML]{C9DAF8}158  \\ \hline
\textbf{Total tiles}      & \multicolumn{5}{c|}{\textbf{88637}} \\ \hline
\end{tabular}}
\end{table*}

\subsubsection{Model Training Strategy and Computational Setup}
\label{sec:HistoCancer_Model_Training}
\label{sec:HistoCancer_Compute}
\mavit{} model was trained using a supervised learning approach based on a patch-based processing strategy. Since WSIs are large, they were divided into non-overlapping image tiles of $512 \times 512$ pixels to facilitate computational efficiency while preserving essential histological features.

The training process involved EfficientNet-B3 as the backbone, initialized with pre-trained ImageNet weights. The convolutional layers from EfficientNet-B3 were transferred and fine-tuned. The SGD optimizer was used with an initial learning rate of \(1 \times 10^{-4}\), a batch size of 8, and an early stopping mechanism based on validation loss to reduce the risk of overfitting. The batch size was set to 8 due to the hardware limitations of the GPU used. The training and evaluation experiments were conducted using an NVIDIA GeForce RTX 3060 GPU, an Intel® Core™ i7-11800U CPU operating at 2.30GHz, and 16 GB of RAM. The model was implemented using Python with TensorFlow-Keras. Data augmentation techniques, including random rotations, flips, and color jittering, were incorporated to enhance model robustness.

Performance metrics included accuracy, sensitivity, specificity, and F1-score, along with calibration curves, confidence intervals, Brier loss, confusion matrices, and heatmap visualizations. These assessments provided a comprehensive analysis of the model's reliability and classification performance.

\begin{table}[h!]
\centering
\caption{Experimental Setups}
\label{tab:HistoCancer_Experiment_setup}
\scalebox{0.65}{
\begin{tabular}{|l|p{8cm}|}
\hline
\textbf{Component}            & \textbf{Details}              \\ \hline
Software                      & Python, Tensorflow, Keras     \\ \hline
GPU                           & NVIDIA GeForce RTX 3060       \\ \hline
CPU                           & Intel® Core™ i7-11800U @ 2.30GHz \\ \hline
RAM                           & 16 GB                         \\ \hline
Optimizer                     & SGD with Momentum             \\ \hline
Evaluation Metrics            & Calibration Curve, Confidence Interval, Brier Loss, Confusion Matrix, Accuracy, Sensitivity, Specificity, F1-Score, Class Prediction Labelmap, Probability Heatmap \\ \hline
\end{tabular}
}
\end{table}

\subsection{Prediction Performance}
\label{sec:HistoCancer_Prediction_Performance}
All reported performance metrics—including sensitivity, specificity, accuracy, and F1-score—are calculated at the \textbf{macro-average} level across the 11 diagnostic classes. This choice reflects an equal weighting of all classes, including less frequent tumor types, and ensures that model performance is not dominated by the majority classes. Diagnosis is obtained by aggregating patch-level results using a majority voting strategy, and evaluation metrics are computed based on these aggregated labels. This macro-level reporting aligns with clinical relevance, as underperforming on minority classes (e.g., rare glioma subtypes) may have significant diagnostic consequences despite their lower prevalence in the dataset.

\begin{figure}[h!]
    \centering
    \includegraphics[width=0.49\textwidth]{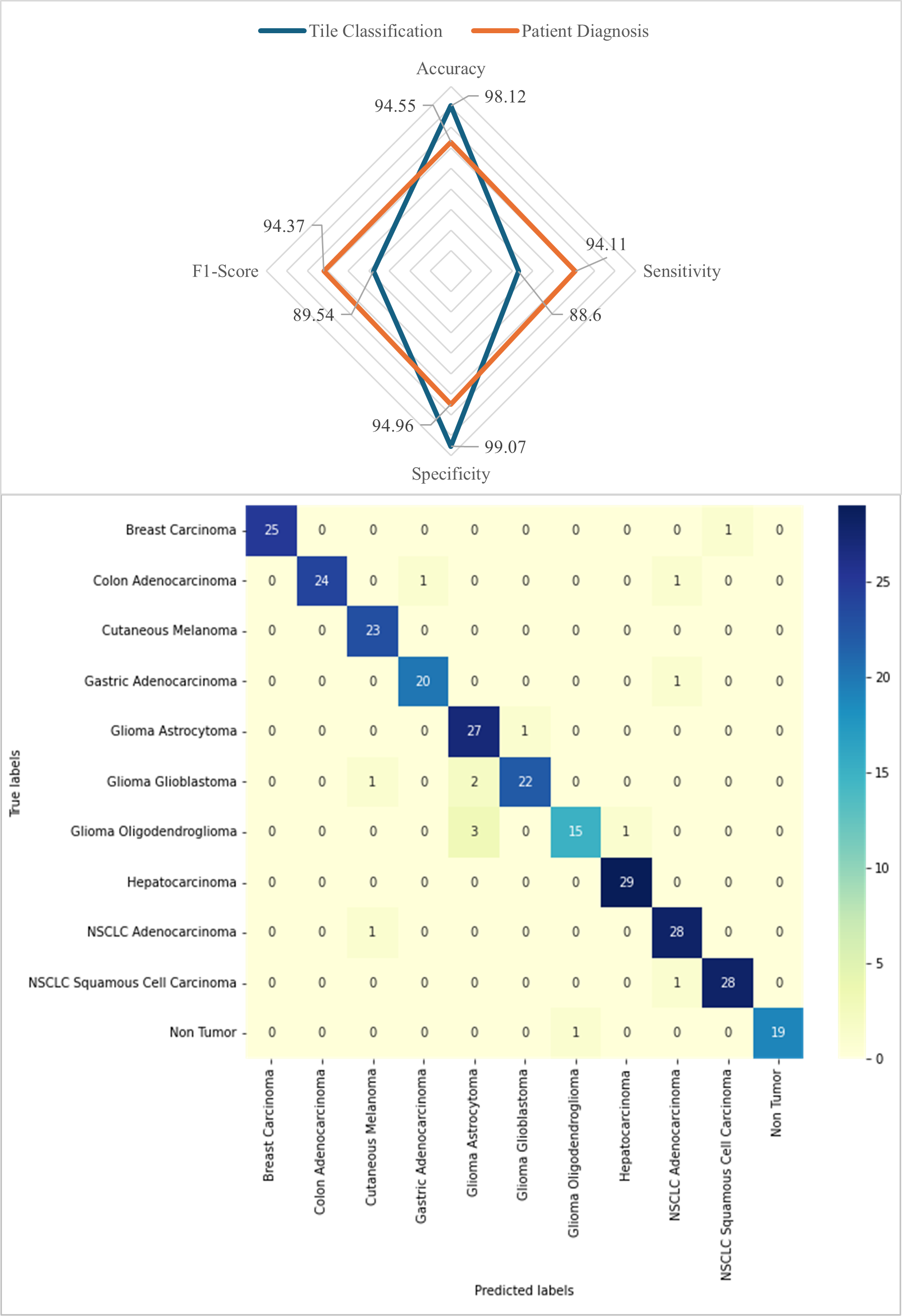}
    \caption{Radar Graph of Accuracy, Sensitivity, Specificity, and F1-Score, with Confusion Matrix Below, for \mavit{} and \method{} on Patient-Level Test Set.}
    \label{fig:1_HistoCancer_Radar_Graph_And_Confusion_Matrix}
\end{figure}

The proposed architecture has been tested on 275 cases. The prediction performance is reported in \autoref{fig:1_HistoCancer_Radar_Graph_And_Confusion_Matrix}. \mavit{} effectively classified 98.12\% of patient patches (Blue curve in Radar graph). In addition, \method{} shows high diagnostic sensitivity (94.11\%), as shown in the orange curve. The confusion matrix provides a detailed breakdown of the model's predictions. As shown, the model correctly identified all cases of Glioma Astrocytoma, Hepatocarcinoma, NSCLC Adenocarcinoma, and NSCLC Squamous Cell Carcinoma. Moreover, the model successfully detects non-tumor WSIs, identifying 19 out of 20 non-tumor cases. This indicates that \method{} was able to distinguish between tumor and non-tumor regions within the evaluated histotypes in the reported cases.

However, it is important to note that there was some confusion between similar gliomas. For example, two cases of Glioma Glioblastoma were misclassified as Glioma Oligodendroglioma and vice versa. This could be attributed to the histopathological similarities in these subtypes of gliomas. Such misclassifications may be related to overlapping visual characteristics within histopathology slides, highlighting the inherent difficulty in differentiating between these closely related cancer types. Despite this, the overall performance underscores the robustness of \method{} in tumor detection and cancer subtype classification.

In this study, we distinguish between two levels of evaluation: \textit{tile-level} and \textit{patient-level}. Patient-level performance refers to the final diagnostic outcome, obtained by aggregating predictions across all WSIs associated with a patient using majority voting, with evaluation metrics computed on a test set split at the patient level. Tile-level performance, on the other hand, represents classification accuracy on individual image tiles extracted from the same patient-based test set—not from a separate or independent split. Hence, both evaluation levels are derived from the same patient-based division, ensuring consistency in validation. Importantly, sensitivity and F1-score serve as the most informative metrics in this context, as they indicate the model's ability to correctly identify cancer cases. As presented in ~\autoref{fig:1_HistoCancer_Radar_Graph_And_Confusion_Matrix}, the tile-level classification yields a sensitivity of 88.60\% and an F1-score of 89.54\%. These metrics further improve at the patient level, reaching 94.11\% and 94.37\%, respectively, underscoring the robustness of the majority-voting aggregation in reducing noise and enhancing clinical relevance.

\subsection{Model Robustness}
\label{sec:HistoCancer_How robust is the model to future inputs?}
\begin{figure}[h!]
    \centering
    \includegraphics[width=0.5\textwidth]{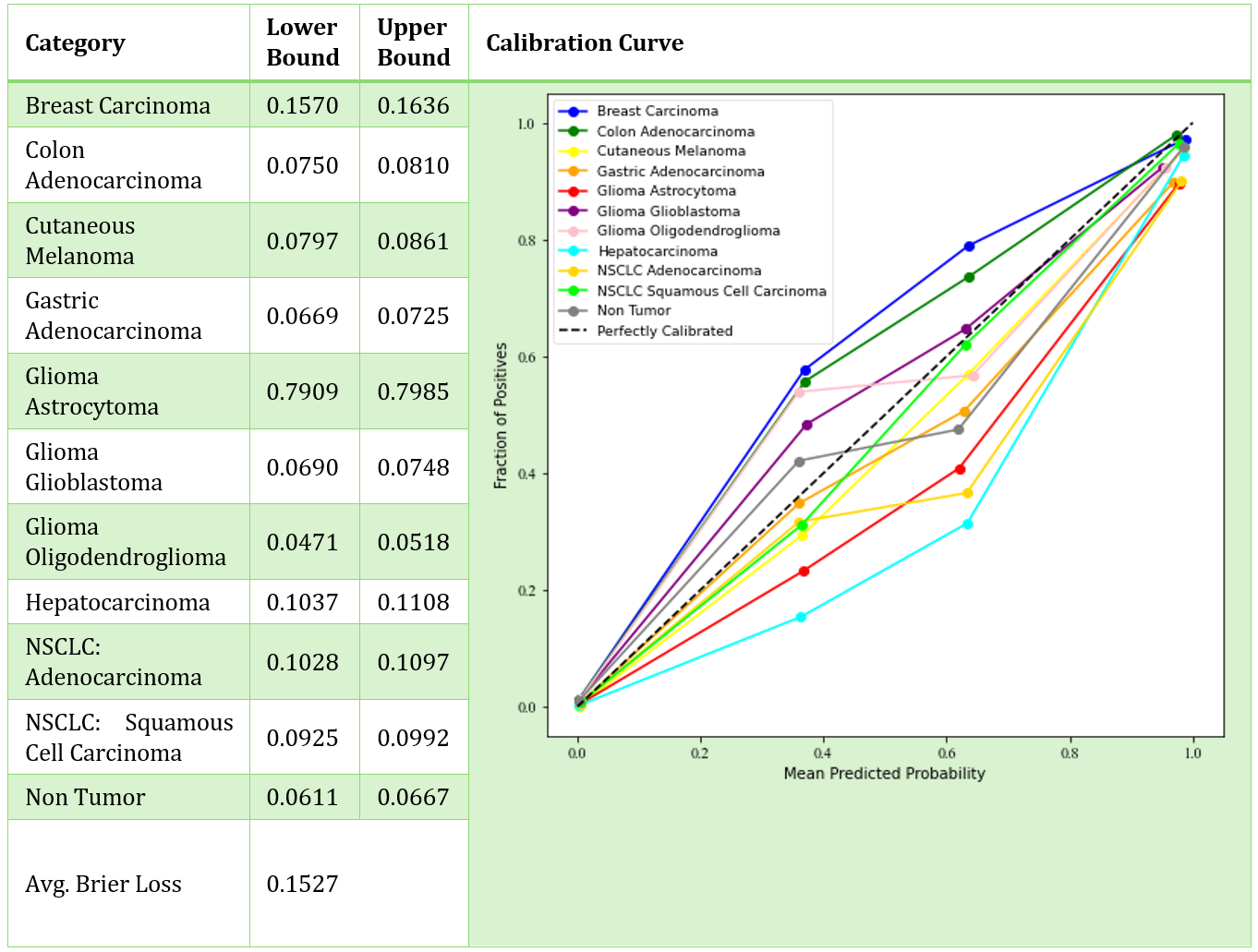}
    \caption{Displays the model's confidence intervals, Brier loss, and calibration curves, which reflect prediction accuracy and reliability. Lower Brier loss (0.1527) indicates better accuracy, while narrower confidence intervals suggest higher confidence. The alignment of calibration curves with the central curve reflects the model's reliability for future predictions.}
    \label{tab:HistoCancer-CAD_Confidence_Intervals}
\end{figure}

As shown in \autoref{tab:HistoCancer-CAD_Confidence_Intervals}, the model calibration curves illustrate the relationship between predicted probabilities and observed results, along with confidence interval values and Brier loss. These results indicate a promising prediction reliability on the test set. The calibration curves on the test set demonstrate how well the predicted probabilities align with the actual outcomes, suggesting that the model is expected to exhibit robustness for future inputs \cite{Austin2020}.

The Brier loss mentioned in this study is used strictly as an \textbf{evaluation metric}—not as a training loss function. It measures the mean squared difference between predicted class probabilities and the true labels, providing insight into the model’s calibration and confidence reliability. During training, we employed the standard categorical cross-entropy loss, as it is more suitable for optimizing multi-class classification tasks. The inclusion of Brier loss in our evaluation offers an additional perspective on the probabilistic accuracy of predictions.

\begin{figure*}[h!]
    \centering
    \includegraphics[width=0.60\textwidth]{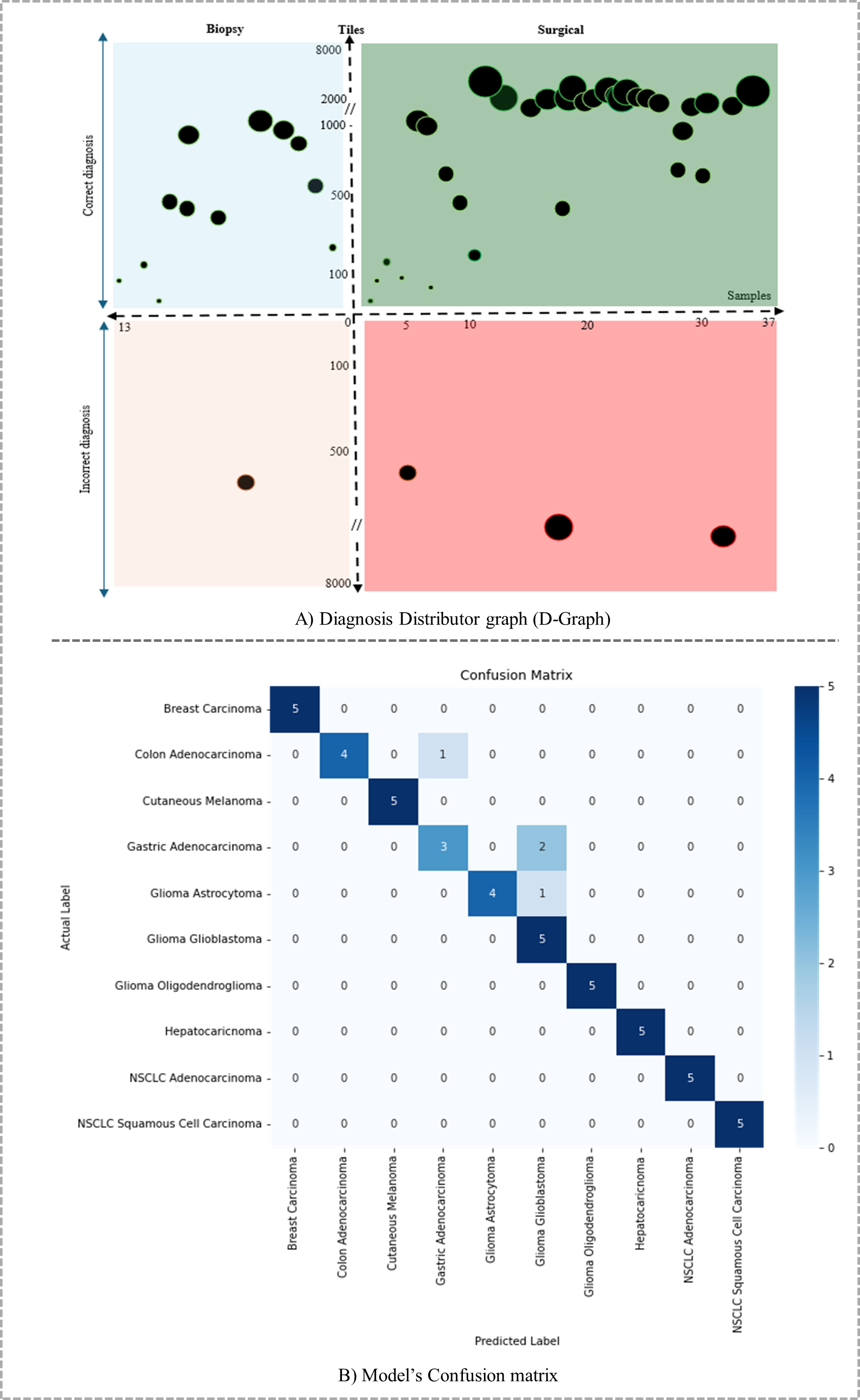}
    \caption{\dgraph{} visualizes the Medical Predictive Model's results (A), with upper quadrants representing correct diagnoses and lower ones indicating incorrect ones. Surgical and biopsy data diagnoses are on the right and left sides of the vertical axis, respectively. Sample counts are on the horizontal axis, and image counts per sample are on the vertical axis. Circle size correlates with image count per sample.}
    \label{fig:2_HistoCancer_D_Graph}
\end{figure*}

\autoref{fig:2_HistoCancer_D_Graph} shows \dgraph{} output for the CliniR-ds dataset. As described in \autoref{sec:HistoCancer_D-Graph}, the graph separates correct and incorrect predictions by quadrant, with surgical and biopsy cases placed to the right and left of the vertical axis, respectively. Each circle represents a sample, with its size corresponding to the number of extracted image tiles. In this evaluation, 46 out of 50 cases were correctly classified, with misclassifications distributed across both modalities.

\autoref{fig:2_HistoCancer_D_Graph} (B) presents the model's confusion matrix derived specifically from the evaluation on the clinical CliniR-ds dataset, which includes both surgical and biopsy specimens. This is distinct from \autoref{fig:1_HistoCancer_Radar_Graph_And_Confusion_Matrix}, which shows the confusion matrix and performance metrics based on the TgCancer-db test set composed solely of TCGA surgical samples. The matrix in \autoref{fig:2_HistoCancer_D_Graph} (B) reflects the model’s generalization capability to unseen, real-world clinical data, while \autoref{fig:1_HistoCancer_Radar_Graph_And_Confusion_Matrix} evaluates performance on curated research data. This distinction underscores \method{}'s robustness and applicability in routine diagnostic settings.

\subsection{Workflow for Patch Generation}
\label{sec:HistoCancer_Workflow for Tile Generation in Experiments}
The patches are generated using SlideTiler \cite{Barcellona2024} in our experimental setup. This step occurs before further analysis with \method{}. To navigate the WSI and effectively generate ROIs using SlideTiler, the following steps are essential:

\begin{enumerate}
    \item \textbf{WSI Loading}: The WSI is loaded into the SlideTiler application, which supports multi-gigabyte image files.
    \item \textbf{Patch Generation}: The pathologist scans the image using the SlideTiler, and marks a WSI or ROI to create a grid of smaller image patches.
\end{enumerate}

The ROI selection workflow allows users to define specific regions for analysis, enabling focused input to \method{} during computational processing.

\subsection{Class Prediction and Probability Heatmap}
\label{sec:HistoCancer_Class prediction and Probability Heatmap}
Visual tools can assist in reviewing and contextualizing model output in image-based diagnostic tasks such as histological interpretation. Two such visualizations—Class Prediction Labelmaps and Probability Heatmaps—are used to represent the model's predictions spatially.

\begin{figure*}[h!]
    \centering
    \includegraphics[width=0.60\textwidth]{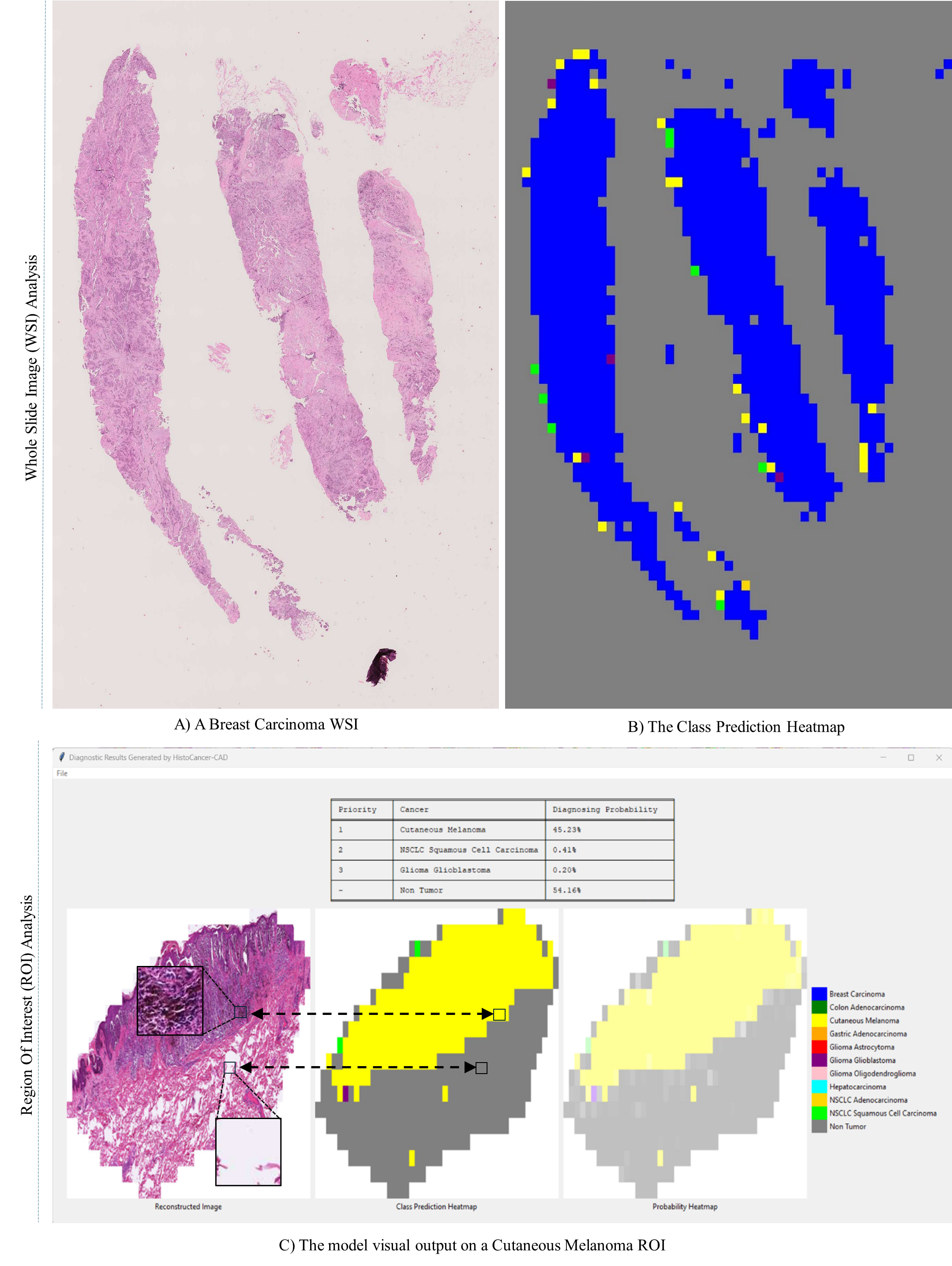}
    \caption{Visual outputs of \method{} processing WSI (part a) and ROI (part c) from two clinical specimen biopsy samples. Initially, the preprocessing steps (described in \ref{sec:HistoCancer_Pre-processing}) segment the WSI into constituent tiles, resulting in 3859 tiles. Subsequent analysis by \method{} generates the visual output shown in (part b). Panel (part c) displays the entire visual output on the \method{} Software GUI.}
    \label{fig:3_HistoCancer_Visual_Output}
\end{figure*}

Class Prediction Labelmaps display the predicted class label for each tile within a WSI or ROI, providing a localized view of the classification results. Probability Heatmaps, on the other hand, indicate the associated confidence values for these predictions, which can highlight areas of high or low certainty. These maps are generated as part of the \method{} output and are shown in \autoref{fig:3_HistoCancer_Visual_Output}. The visual examples illustrate how the model processes both user-defined ROIs and entire WSIs. In particular, \autoref{fig:3_HistoCancer_Visual_Output} (part c) shows regions labeled as tumor and non-tumor within the same tissue section, as predicted by the model.

\subsection{Model’s Time Complexity}
\label{sec:HistoCancer_Model’s Time Complexity}
The time complexity of \method{} was evaluated using the hardware configuration listed in \autoref{tab:HistoCancer_Experiment_setup}. The architecture consists of 390 layers, with 14794931 trainable parameters and 74311 non-trainable parameters.

During training, the average processing time was measured at 0.0365 seconds per image patch. For inference, the prediction time per patch decreased to 0.0183 seconds. The WSI reconstruction process required 0.0168 seconds per tile. Additionally, the generation of class prediction Labelmaps and probability Heatmaps took 0.0035 seconds and 0.0233 seconds per tile, respectively.

\subsection{Ablation Study}
\label{sec:Ablation_Study}
To evaluate the contribution of each component within the proposed method, an ablation study was conducted by incrementally adding components to the baseline EfficientNet-B3 architecture. The comparison includes:

\begin{itemize}
    \item \textbf{EfficientNet-B3 (Baseline)}: The standard CNN backbone trained from ImageNet pre-trained weights with all layers unfrozen.
    \item \textbf{EfficientNet-B3 + VTM}: Adding the VTM to incorporate global contextual information.
    \item \textbf{\mavit{} (EfficientNet-B3 + VTM + DFS)}: Full architecture including both the VTM and DFS.
\end{itemize}

As presented in \autoref{tab:ablation}, the baseline EfficientNet-B3 performs strongly with a high Specificity of 99.86\%, but comparatively lower Sensitivity. Adding the VTM improves Sensitivity and F1-score, while slightly lowering specificity. This change reflects a more balanced classification behavior. Integrating the DFS in the final \mavit{} model improves all metrics, particularly in Sensitivity and F1-score. This suggests that combining the multi-level feature fusion and global context representation enhances the model’s discriminative performance.

\begin{table}[h!]
\centering
\caption{Ablation study results on the TgCancer-db test set (Acc: Accuracy, Sen: Sensitivity, Spec: Specificity)}.
\label{tab:ablation}
\scalebox{0.75}{
\begin{tabular}{|l|c|c|c|c|}
\hline
\textbf{Model Variant} & \textbf{Acc} & \textbf{Sen} & \textbf{Spec} & \textbf{F1-Score} \\
\hline
EfficientNet-B3 & 97.36\% & 85.01\% & \textbf{99.86\%} & 86.71\% \\
\hline
+ VTM & 97.65\% & 87.36\% & 99.55\% & 87.91\% \\
\hline
+ DFS (\textbf{\mavit{}}) & \textbf{98.12\%} & \textbf{88.60\%} & 99.07\% & \textbf{89.54\%} \\
\hline
\end{tabular}
}
\end{table}

\section{Discussion}
\label{sec:HistoCancer_Discussion}
Integrating AI models with pathology knowledge remains essential to advance DP. The \method{} was designed to support the classification and diagnosis of 10 tumor histotypes and non-tumor areas. The system integrates multiple architectural elements introduced in this work: \mavit{} (\autoref{sec:HistoCancer_ImageClassification_MAViT}), VTM (\autoref{sec:HistoCancer_VTM}), DFS (\autoref{sec:HistoCancer_DFS}), and \dgraph{} (\autoref{fig:2_HistoCancer_D_Graph}). \mavit{} is a Vision Transformer designed specifically for histopathology image classification. Its modular structure supports the future integration of additional tumor types and diagnostic modules.

To address challenges in hierarchical feature representation, DFS employs Early and Late Fusion, enabling feature integration across different levels of abstraction. Early Fusion combines intermediate representations, while Late Fusion aggregates refined outputs prior to transformer-based attention. This dual-level fusion supports feature processing across scales \cite{Nagrani2021, Singh2022, Mauricio2023}.

A comparative evaluation is presented in \autoref{tab:HistoCancer_MAViT_Comparison}, benchmarking \mavit{} against widely used models including MobileNetV3Large~\cite{Howard2019}, ResNet50~\cite{He2016}, EfficientNet-B3~\cite{Tan2019}, and ViT~\cite{Dosovitskiy2020}. All models were re-trained and evaluated on the TgCancer-db dataset under identical conditions. \mavit{} achieved the highest reported accuracy (98.12\%), along with the top sensitivity, specificity, and F1-score. These results are based on predictions aggregated for cancer diagnosis using majority voting, as illustrated in \autoref{fig:1_HistoCancer_Radar_Graph_And_Confusion_Matrix}. 
While architectural innovation alone may yield marginal gains in benchmark accuracy, our design of \method{} prioritizes deployability, flexibility, and interpretability in clinical pathology settings—objectives not sufficiently addressed by existing models. As shown in ~\autoref{tab:HistoCancer_MAViT_Comparison}, despite \mavit{}'s strong performance, \method{} framework effectively hosts various CNN and transformer models, including MobileNetV3, ResNet50, and ViT-B16. This not only emphasizes architectural neutrality but also underscores our system’s adaptability to different constraints—whether speed, model size, or diagnostic focus. The full ablation study (\autoref{tab:ablation}) further isolates the contributions of each major module: VTM for contextual awareness, and DFS for multi-scale feature integration. Unlike black-box designs, \method{} also integrates a structured visualization toolkit—including class labelmaps, probability heatmaps, and the new D-Graph (\autoref{fig:2_HistoCancer_D_Graph})—which moves beyond qualitative saliency and supports measurable insight into model behavior across biopsies and surgical samples. Therefore, the technical emphasis in our framework is not solely architectural; rather, it represents a system-level approach that combines performance, transparency, and operational versatility for real-world diagnostic support.

\begin{table*}[h!]
\centering
\caption{Comparison of \mavit{} with CNNs within the \method{} framework on TgCancer-db in terms of classification performance and computational complexity. This table demonstrates that, despite the advantages of \mavit{}, \method{} is a strong and flexible host framework for integrating various CNNs for cancer diagnosis, accommodating diverse needs such as lightweight architectures, deeper models, or efficiency-oriented designs.}

\label{tab:HistoCancer_MAViT_Comparison}
\scalebox{0.85}{
\begin{tabular}{|l|c|c|c|c|c|c|}
\hline
\textbf{Model} & \textbf{Acc} & \textbf{Sen} & \textbf{Spec} & \textbf{F1-Score} & \textbf{Parameters} & \textbf{Time (ms)} \\
\hline
MobileNetV3L \cite{Howard2019} & 97.76 & 86.55 & 98.88 & 87.53 & 4,240,523 & 11.50 \\
\hline
ResNet50 \cite{He2016} & 97.59 & 86.27 & 98.72 & 86.67 & 23,610,251 & 17.60 \\
\hline
EfficientNet-B3 \cite{Tan2019} & 97.36 & 85.01 & 98.89 & 86.71 & 10,800,442 & 32.60 \\
\hline
ViT\_B16 \cite{Dosovitskiy2020} & 97.94 & 87.64 & 98.97 & 88.57 & 147,949,931 & 37.30 \\
\hline
\textbf{\mavit{}} & \textbf{98.12} & \textbf{88.60} & \textbf{99.07} & \textbf{89.54} & \textbf{148,692,942} & \textbf{37.70} \\
\hline
\end{tabular}
}
\end{table*}

\begin{figure}[h!]
\centering
\includegraphics[width=0.5\textwidth]{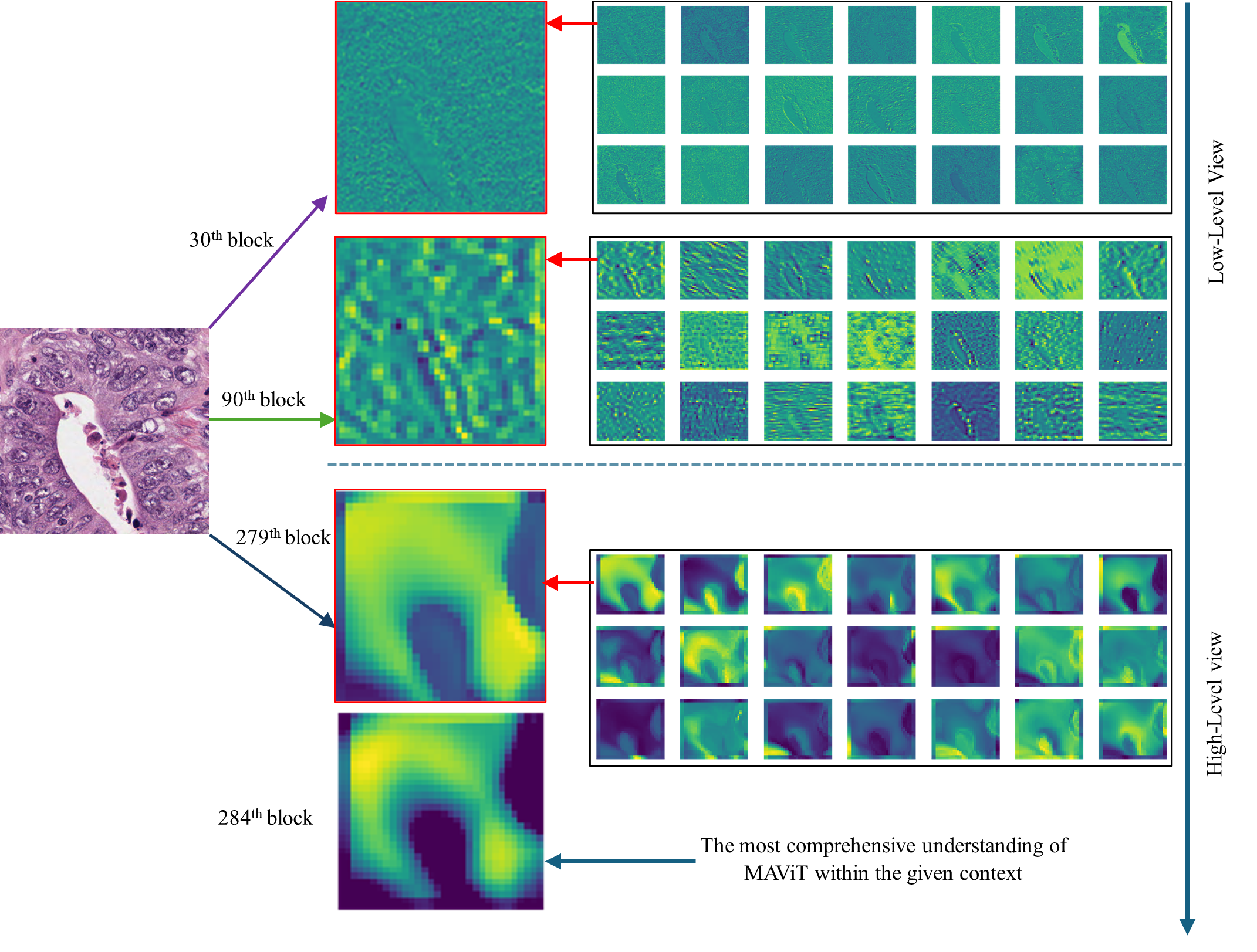}
\caption{Class Activation Maps from blocks 279 and 284 showing attention to key regions in histopathology slides.}
\label{fig:12_HistoCancer_MAViT_CAM}
\end{figure}

\autoref{fig:12_HistoCancer_MAViT_CAM} presents Class Activation Maps that show patch-level feature relevance at different blocks within \mavit{}. These results support the model’s capacity to extract local contextual cues within WSIs.

As discussed in \autoref{sec:HistoCancer_Model’s Time Complexity}, the training time per patch is 37.70 ms. For 358,589 training tiles, one epoch requires approximately 225 minutes.

The non-tumor (healthy) tissue was explicitly included as a class during both training and evaluation. In addition, the dataset includes distinct subtypes of brain gliomas—Glioblastoma, Astrocytoma, and Oligodendroglioma—as well as two lung cancer subtypes: Lung Adenocarcinoma and Lung Squamous Cell Carcinoma. Furthermore, Gastric Adenocarcinoma and Colon Adenocarcinoma were also included, both originating from the digestive system and may share similar histological patterns. This strategy for selecting diverse tumor subtypes supports the model’s ability to learn discriminative features beyond simple organ-level cues.

\begin{figure*}[h!]
\centering
\includegraphics[width=0.85\textwidth]{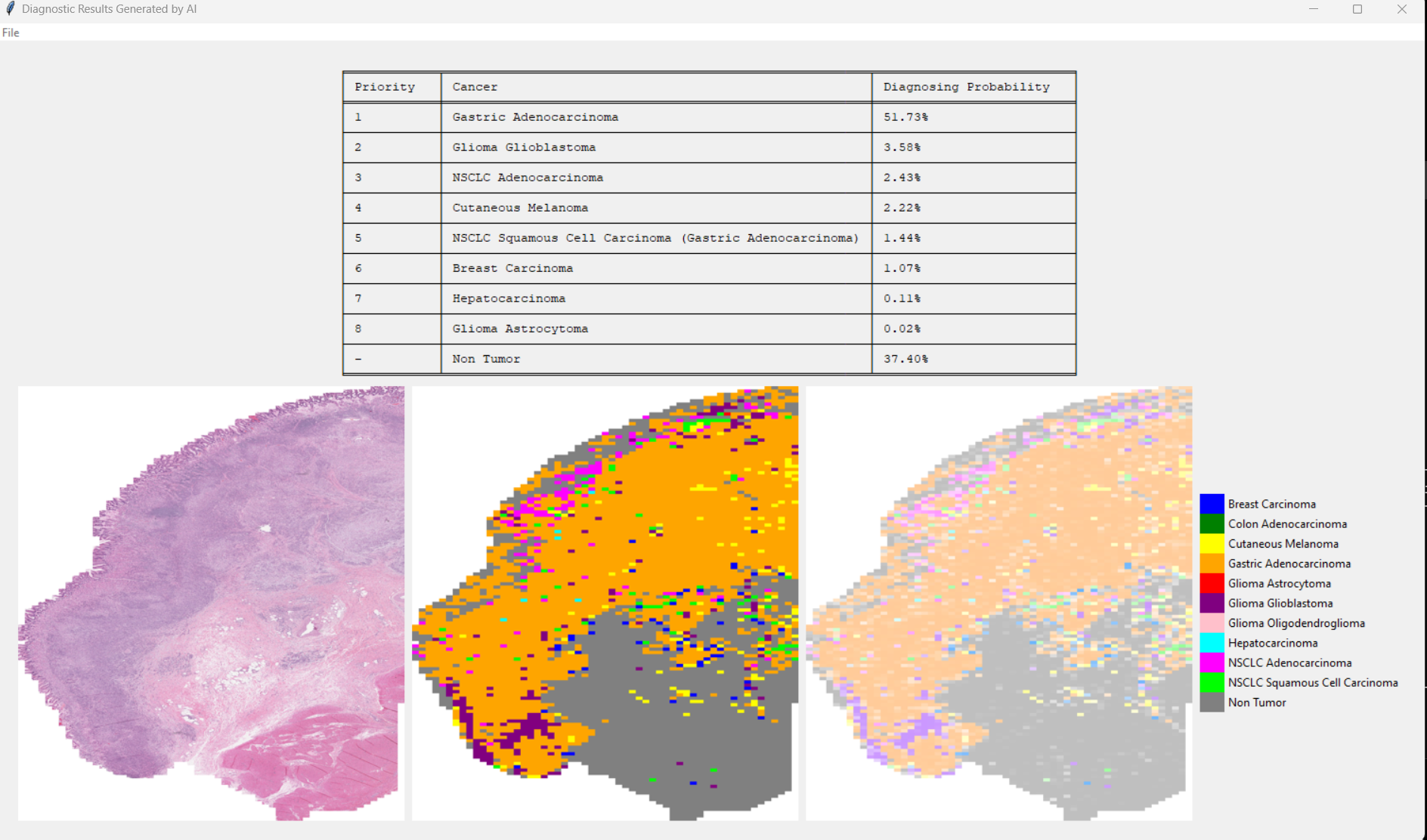}
\caption{Example of tumor and non-tumor differentiation within the same ROI in gastric carcinoma.}
\label{fig:Gastric}
\end{figure*}

The \method{} framework is designed with operational flexibility to support both WSI and ROI modes, accommodating a variety of diagnostic workflows. In WSI mode, the model automatically analyzes the entire tissue landscape, enabling the detection and classification of tumor and non-tumor regions without requiring manual input. This mode is particularly suited for initial screening or identifying multifocal pathology. ROI mode, on the other hand, allows a pathologist to target specific areas—such as ambiguous glioma regions—for focused analysis, thereby optimizing computational efficiency when full-slide evaluation is unnecessary. Notably, ROI selection is entirely optional and does not constrain the model’s capabilities. In both modes, the system can identify heterogeneous tissue regions, including tumor and adjacent non-tumor areas, as evidenced by the visualizations in \autoref{fig:Gastric} and \autoref{fig:3_HistoCancer_Visual_Output}. These examples underscore that the model captures tumor-specific histomorphological features rather than relying on organ-level context, reinforcing its alignment with the real diagnostic challenges encountered in clinical pathology.

It is important to note that the model is trained on a fixed number of 100 randomly selected patches per WSI (Not all regions of the WSI). As a result, the model never observes the whole tissue architecture of an entire organ during training, which reduces the likelihood of learning global organ-specific patterns. While some degree of organ-related features may still be present on patches, selecting a few shots of WSI's patches during training—combined with randomized sampling—further encourages the model to focus on tumor-specific features rather than organ morphology. Regarding the misclassification between brain glioma tumor subtypes, this is indeed a recognized challenge in pathology, as glioma subtypes often share overlapping histological features, making the prediction difficult even under manual pathological assessment. Nevertheless, the observed confusion is limited to specific brain tumor categories (e.g., Astrocytoma vs. Oligodendroglioma) and does not extend to cross-organ misclassification within the test set. Furthermore, in the confusion matrix (part B) of \autoref{fig:2_HistoCancer_D_Graph}, which corresponds to real clinical cases from Ospedale Dell’Angelo in Mestre-Venice, the model achieved 14 correct predictions out of 15 cases, with 5 cases each for Glioblastoma, Oligodendroglioma, and Astrocytoma. This supports the interpretation that the misclassifications observed in other parts of the evaluation are not specifically due to the model relying on organ-specific cues but rather may reflect the inherent diagnostic difficulty associated with these particular glioma subtypes.

Although organ-level classification may appear trivial in certain diagnostic settings, the classification task presented in this study involves a much finer granularity. \method{} was trained and evaluated on several histologically similar tumor subtypes within the same organ systems, such as glioblastoma, astrocytoma, and oligodendroglioma for the brain; adenocarcinoma and squamous cell carcinoma for non-small cell lung cancer; and colon versus gastric adenocarcinoma within the gastrointestinal tract. These distinctions are well-recognized challenges even for expert pathologists due to overlapping histomorphological features. Moreover, by explicitly including early-stage invasive carcinomas and non-tumor tissues within the classification framework, the model is equipped to support nuanced diagnostic decisions reflective of routine clinical practice. The diversity and complexity of the included classes ensure that the model does not simply perform high-level tissue recognition but instead learns discriminative, pathology-relevant features required for real-world diagnostic assistance.

\autoref{tab:sota_comparison} summarizes the reported performance of several recent models developed for cancer diagnosis using histopathology images. These include MI-Zero \cite{lu2023mi-zero}, KEEP \cite{zhou2024keep}, HIPT \cite{chen2022HIPT}, and Pan-Cancer MMF \cite{chen2022pancancer}. While evaluation settings and datasets may differ, the comparison offers a contextual overview of classification outcomes across common metrics.

TOAD, which focuses on tumor origin prediction, reports an accuracy of 84\% in internal evaluations. MI-Zero—a zero-shot cancer subtyping approach using vision-language pretraining—achieves a median accuracy of 70.2\% across three subtypes. KEEP emphasizes knowledge-guided vision-language alignment and reports high sensitivity (89.8\%) and specificity (95.0\%) across seven cancer types. HIPT, which applies a hierarchical transformer architecture, shows a high F1-score of 0.934 on lung adenocarcinoma (LUAD) classification, while Pan-Cancer MMF adopts a multimodal strategy integrating genomics and pathology, reporting a c-index of 0.644 across 14 cancer types. Compared to these models, \method{}, in a supervised multi-class classification setting, reports an accuracy of 94.55\%, a sensitivity of 94.11\%, a specificity of 94.96\%, and an F1-score of 94.37\%. Although direct comparisons should be interpreted cautiously due to differences in datasets and experimental protocols, these values suggest that \method{} performs competitively in conventional diagnostic metrics.

\begin{table}[h!]
\centering
\caption{\method{} comparison with state-of-the-art models for cancer diagnosis.}
\label{tab:sota_comparison}
\scalebox{0.7}{
\begin{tabular}{|l|c|c|c|c|}
\hline
\textbf{Model} & \textbf{Acc} & \textbf{Sen} & \textbf{Spec} & \textbf{F1-Score} \\
\hline
TOAD~\cite{Lu2021} & 84\% & -- & -- & -- \\
\hline
MI-Zero~\cite{lu2023mi-zero} & 70.20\% & -- & -- & -- \\
\hline
KEEP~\cite{zhou2024keep} & -- & 89.80\% & \textbf{95\%} & -- \\
\hline
HIPT~\cite{chen2022HIPT} & -- & -- & -- & 0.934 (LUAD) \\
\hline
Pan-Cancer MMF~\cite{chen2022pancancer} & -- & -- & -- & c-Index: 0.644 \\
\hline
\textbf{\method{}} & \textbf{94.55\%} & \textbf{94.11\%} & 94.96\% & \textbf{94.37\%} \\
\hline
\end{tabular}
}
\end{table}

While the classification of multiple tumor types has been explored in previous works, including the study by ~\cite{lu2021aipathology} on cancers of unknown primary origin, the objectives, data structure, and clinical use cases of our study differ substantially. The primary aim of Lu et al.'s work is to infer the likely site of origin for metastatic tumors when the primary site is unknown—addressing a tumor-origin prediction task rooted in genomic and histological integration. In contrast, \method{} is designed as a clinically deployable, end-to-end CAD system for \textit{direct histopathological diagnosis} of primary tumor subtypes across multiple organ systems using H\&E-stained WSIs. Our model focuses on \textit{subtype-level classification} of diagnostically overlapping cancers, including glioma subtypes and non-small cell lung cancer variants, which present significant morphological ambiguity under routine histology. Moreover, we present an architecture tailored for high-resolution WSI analysis, validated on both a large multi-class cohort (TgCancer-ds) and a real-world clinical dataset (CliniR-ds), including biopsy samples—a scenario not addressed in the referenced work. The task novelty thus lies not only in the diversity of cancer types considered but also in the clinically grounded design, subtype resolution, support for both WSI- and ROI-based diagnostics, and a visual and probabilistic interface supporting pathologist trust. These components, taken together, demonstrate that our work does not replicate but meaningfully extends the current landscape of AI in digital pathology.

While widely-used public benchmarks such as Camelyon16/17 and CRC datasets \cite{bejnordi2017camelyon16,bandi2018detection,kather2019predicting} have played a bold role in the development of digital pathology tools, our study prioritizes clinically grounded validation over benchmark-oriented evaluation. \method{} was developed and rigorously tested on two diverse cohorts: the TgCancer-ds dataset, sourced from TCGA and encompassing ten distinct tumor types and non-tumor tissue, and the CliniR-ds dataset, a real-world clinical collection of biopsy and surgical cases from routine pathology practice. This validation strategy reflects the practical diagnostic complexity encountered in clinical workflows and offers broader histological diversity than single-entity datasets. By focusing on multi-class tumor subtype recognition in real-world whole-slide images, our evaluation aims to ensure clinical applicability and generalizability across heterogeneous tissue presentations, which are not adequately captured by binary classification tasks such as tumor detection in Camelyon16 or CRC.

\section{Conclusions}
\label{sec:HistoCancer_Conclusions}
\method{} presents a modular and deployable framework for multi-class tumor classification in histopathological whole slide images, integrating a vision transformer architecture, hierarchical feature fusion, and interpretability tools. The system is designed to support clinical decision-making by combining computational accuracy with visual transparency. Comprehensive evaluation across internal (TgCancer-db) and external (CliniR-ds) datasets demonstrates its robust diagnostic performance and favorable computational efficiency across a broad spectrum of tumor subtypes. A current limitation lies in the relatively limited representation of biopsy-derived samples in the training set, which may affect generalization in certain clinical scenarios. Future work will explore the inclusion of larger and more diverse biopsy cohorts, as well as the incorporation of soft-label aggregation mechanisms to address classification ambiguity in morphologically heterogeneous regions. To facilitate reproducibility and community-driven advancement, \method{} is publicly released for research use, providing a foundation for further innovation in AI-powered digital pathology.

\section*{Data Availability}
\label{sec:HistoCancer_Data_Availability}

The software and source code developed for this study will be publicly available via the author's \href{https://github.com/AshkanShakarami/DepViT-CAD/tree/main}{GitHub} repository. The WSI used for making TgCancer-ds have been acquired from the GDC portal at \url{https://gdc-portal.nci.nih.gov} and includes cases from multiple TCGA cohorts: lung adenocarcinoma (TCGA-LUAD), lung squamous cell carcinoma (TCGA-LUSC), breast carcinoma (TCGA-BRCA), hepatic carcinoma (TCGA-LIHC), colon adenocarcinoma (TCGA-COAD), gastric adenocarcinoma (TCGA-STAD), melanoma (TCGA-SKCM), low-grade gliomas (TCGA-LGG, including oligodendroglioma and astrocytoma), and glioblastoma (TCGA-GBM). The list of WSIs used in TgCancer-ds will be made available via the author's \href{https://github.com/AshkanShakarami/DepViT-CAD/tree/main}{GitHub}.

The CliniR-ds dataset, which comprises real-world clinical cases from the Angelo Hospital Pathology Unit in Mestre, Venice, is not publicly available due to patient privacy and institutional restrictions.

\bibliographystyle{unsrt}
\bibliography{references}

\begin{thebibliography}{10}

\bibitem{Rosai2007}
J.~Rosai.
\newblock Why microscopy will remain a cornerstone of surgical pathology.
\newblock {\em Laboratory Investigation}, 87(5):403--408, 2007.

\bibitem{Pisapia2022}
P.~Pisapia, V.~L'Imperio, F.~Galuppini, E.~Sajjadi, A.~Russo, B.~Cerbelli, and U.~Malapelle.
\newblock The evolving landscape of anatomic pathology.
\newblock {\em Critical Reviews in Oncology/Hematology}, 178:103776, 2022.

\bibitem{Metter2019}
D.~M. Metter, T.~J. Colgan, S.~T. Leung, C.~F. Timmons, and J.~Y. Park.
\newblock Trends in the us and canadian pathologist workforces from 2007 to 2017.
\newblock {\em JAMA Network Open}, 2(5):e194337--e194337, 2019.

\bibitem{Markl2021}
B.~Märkl, L.~Füzesi, R.~Huss, S.~Bauer, and T.~Schaller.
\newblock Number of pathologists in germany: Comparison with european countries, usa, and canada.
\newblock {\em Virchows Archiv}, 478:335--341, 2021.

\bibitem{Betmouni2021}
S.~Betmouni.
\newblock Diagnostic digital pathology implementation: Learning from the digital health experience.
\newblock {\em Digital Health}, 7:20552076211020240, 2021.

\bibitem{Hanna2022}
M.~G. Hanna, O.~Ardon, V.~E. Reuter, S.~J. Sirintrapun, C.~England, D.~S. Klimstra, and M.~R. Hameed.
\newblock Integrating digital pathology into clinical practice.
\newblock {\em Modern Pathology}, 35(2):152--164, 2022.

\bibitem{Maier2019}
A.~Maier, C.~Syben, T.~Lasser, and C.~Riess.
\newblock A gentle introduction to deep learning in medical image processing.
\newblock {\em Zeitschrift für Medizinische Physik}, 29(2):86--101, 2019.

\bibitem{Echle2021}
A.~Echle, N.~T. Rindtorff, T.~J. Brinker, T.~Luedde, A.~T. Pearson, and J.~N. Kather.
\newblock Deep learning in cancer pathology: A new generation of clinical biomarkers.
\newblock {\em British Journal of Cancer}, 124(4):686--696, 2021.

\bibitem{Song2023}
A.~H. Song, G.~Jaume, D.~F. Williamson, M.~Y. Lu, A.~Vaidya, T.~R. Miller, and F.~Mahmood.
\newblock Artificial intelligence for digital and computational pathology.
\newblock {\em Nature Reviews Bioengineering}, 1(12):930--949, 2023.

\bibitem{Baxi2022}
V.~Baxi, R.~Edwards, M.~Montalto, and S.~Saha.
\newblock Digital pathology and artificial intelligence in translational medicine and clinical practice.
\newblock {\em Modern Pathology}, 35(1):23--32, 2022.

\bibitem{VanDiest2024}
P.~J. van Diest, R.~N. Flach, C.~van Dooijeweert, S.~Makineli, G.~E. Breimer, N.~Stathonikos, and M.~Veta.
\newblock Pros and cons of artificial intelligence implementation in diagnostic pathology.
\newblock {\em Histopathology}, 2024.

\bibitem{Shakarami2025thesis}
A.~Shakarami.
\newblock Transformative ai for automating histopathology workflows.
\newblock {\em University of Padova}, 2025.

\bibitem{oh2021pathcnn}
J.~H. Oh, W.~Choi, E.~Ko, M.~Kang, A.~Tannenbaum, and J.~O. Deasy.
\newblock Pathcnn: interpretable convolutional neural networks for survival prediction and pathway analysis applied to glioblastoma.
\newblock {\em Bioinformatics}, 37(Supplement\_1):i443--i450, 2021.

\bibitem{Tan2019}
M.~Tan and Q.~Le.
\newblock Efficientnet: Rethinking model scaling for convolutional neural networks.
\newblock In {\em International Conference on Machine Learning}, pages 6105--6114. PMLR, 2019.

\bibitem{Shakarami2023}
A.~Shakarami, L.~Nicolè, M.~Terreran, A.~P. Dei~Tos, and S.~Ghidoni.
\newblock Tcnn: A transformer convolutional neural network for artifact classification in whole slide images.
\newblock {\em Biomedical Signal Processing and Control}, 84:104812, 2023.

\bibitem{shakarami2024histo}
A.~Shakarami, L.~Nicole, and S.~Ghidoni.
\newblock Ai for advanced cancer diagnosis: a cad system empowered by a novel vision transformer network for histopathology analysis.
\newblock {\em Virchows Archiv}, 485(S1):S397--S398, 2024.

\bibitem{Shakarami2020a}
A.~Shakarami, H.~Tarrah, and A.~Mahdavi-Hormat.
\newblock A cad system for diagnosing alzheimer’s disease using 2d slices and an improved alexnet-svm method.
\newblock {\em Optik}, 212:164237, 2020.

\bibitem{Shakarami2021b}
A.~Shakarami, M.~B. Menhaj, and H.~Tarrah.
\newblock Diagnosing covid-19 disease using an efficient cad system.
\newblock {\em Optik}, 241:167199, 2021.

\bibitem{shakarami2021yolov3}
A.~Shakarami, M.~B. Menhaj, A.~Mahdavi-Hormat, and H.~Tarrah.
\newblock A fast and yet efficient yolov3 for blood cell detection.
\newblock {\em Biomedical Signal Processing and Control}, 66:102495, 2021.

\bibitem{Lu2021}
Ming~Y. Lu, Drew F.~K. Williamson, Tzu-Ming Chen, and et~al.
\newblock Data-efficient and weakly supervised computational pathology on whole-slide images.
\newblock {\em Nature Biomedical Engineering}, 5(6):555--570, 2021.

\bibitem{Dosovitskiy2020}
A.~Dosovitskiy, L.~Beyer, A.~Kolesnikov, D.~Weissenborn, X.~Zhai, T.~Unterthiner, and N.~Houlsby.
\newblock An image is worth 16x16 words: Transformers for image recognition at scale.
\newblock {\em arXiv preprint arXiv:2010.11929}, 2020.

\bibitem{Touvron2021}
H.~Touvron, M.~Cord, M.~Douze, F.~Massa, A.~Sablayrolles, and H.~Jégou.
\newblock Training data-efficient image transformers \& distillation through attention.
\newblock In {\em International Conference on Machine Learning}, pages 10347--10357. PMLR, 2021.

\bibitem{liu2021swin}
Z.~Liu, Y.~Lin, Y.~Cao, H.~Hu, Y.~Wei, Z.~Zhang, S.~Lin, and B.~Guo.
\newblock Swin transformer: Hierarchical vision transformer using shifted windows.
\newblock {\em Proceedings of the IEEE Conference on Computer Vision and Pattern Recognition}, 2021.

\bibitem{Shao2021TransMIL}
Zhihao Shao, Heng Bian, Yicheng Chen, Yingying Wang, Jun Zhang, and Xiangyang Ji.
\newblock Transmil: Transformer based correlated multiple instance learning for whole slide image classification.
\newblock In {\em Advances in Neural Information Processing Systems (NeurIPS)}, volume~34, pages 2136--2147, 2021.

\bibitem{chen2022HIPT}
R.~J. Chen, C.~Chen, Y.~Li, T.~Y. Chen, A.~D. Trister, R.~G. Krishnan, and F.~Mahmood.
\newblock Scaling vision transformers to gigapixel images via hierarchical self-supervised learning.
\newblock In {\em Proceedings of the IEEE/CVF Conference on Computer Vision and Pattern Recognition}, pages 16144--16155, 2022.

\bibitem{Li2023}
G.~Li, G.~Wu, G.~Xu, C.~Li, Z.~Zhu, Y.~Ye, and H.~Zhang.
\newblock Pathological image classification via embedded fusion mutual learning.
\newblock {\em Biomedical Signal Processing and Control}, 79:104181, 2023.

\bibitem{Chen2023}
R.~J. Chen, T.~Ding, M.~Y. Lu, D.~F. Williamson, G.~Jaume, B.~Chen, and F.~Mahmood.
\newblock A general-purpose self-supervised model for computational pathology.
\newblock {\em arXiv preprint arXiv:2308.15474}, 2023.

\bibitem{CONCH}
M.~Y. Lu, B.~Chen, D.~F. Williamson, R.~J. Chen, I.~Liang, T.~Ding, and F.~Mahmood.
\newblock A visual-language foundation model for computational pathology.
\newblock {\em Nature Medicine}, 30(3):863--874, 2024.

\bibitem{mSTAR}
Y.~Xu, Y.~Wang, F.~Zhou, J.~Ma, S.~Yang, H.~Lin, and H.~Chen.
\newblock A multimodal knowledge-enhanced whole-slide pathology foundation model.
\newblock {\em arXiv preprint arXiv:2407.15362}, 2024.

\bibitem{SimCLR}
Ting Chen, Simon Kornblith, Mohammad Norouzi, and Geoffrey Hinton.
\newblock A simple framework for contrastive learning of visual representations.
\newblock In {\em International Conference on Machine Learning}, pages 1597--1607. PMLR, 2020.

\bibitem{MoCo}
Kaiming He, Haoqi Fan, Yuxin Wu, Saining Xie, and Ross Girshick.
\newblock Momentum contrast for unsupervised visual representation learning.
\newblock In {\em Proceedings of the IEEE/CVF Conference on Computer Vision and Pattern Recognition}, pages 9729--9738, 2020.

\bibitem{chen2020pathomic}
R.~J. Chen, M.~Y. Lu, J.~Wang, D.~F. Williamson, S.~J. Rodig, N.~I. Lindeman, and F.~Mahmood.
\newblock Pathomic fusion: an integrated framework for fusing histopathology and genomic features for cancer diagnosis and prognosis.
\newblock {\em IEEE Transactions on Medical Imaging}, 41(4):757--770, 2022.

\bibitem{Barcellona2024}
L.~Barcellona, L.~Nicolè, R.~Cappellesso, A.~P. Dei~Tos, and S.~Ghidoni.
\newblock Slidetiler: A dataset creator software for boosting deep learning on histological whole slide images.
\newblock {\em Journal of Pathology Informatics}, 15:100356, 2024.

\bibitem{Linformer2020}
S.~Wang, B.~Z. Li, M.~Khabsa, H.~Fang, and H.~Ma.
\newblock Linformer: Self-attention with linear complexity.
\newblock {\em arXiv preprint}, arXiv:2006.04768, 2020.

\bibitem{huang2024learnable}
Y.~Y. Huang and W.~T. Chu.
\newblock Learnable context in multiple instance learning for whole slide image classification and segmentation.
\newblock {\em Journal of Imaging Informatics in Medicine}, 1:1--15, 2024.

\bibitem{couture2018multiple}
H.~D. Couture, J.~S. Marron, C.~M. Perou, M.~A. Troester, and M.~Niethammer.
\newblock Multiple instance learning for heterogeneous images: Training a cnn for histopathology.
\newblock In {\em Medical Image Computing and Computer Assisted Intervention–MICCAI 2018: 21st International Conference, Granada, Spain, September 16-20, 2018, Proceedings, Part II}, volume 11071 of {\em Lecture Notes in Computer Science}, pages 254--262. Springer International Publishing, 2018.

\bibitem{Collins2015}
G.~S. Collins, J.~B. Reitsma, D.~G. Altman, and K.~G. Moons.
\newblock Transparent reporting of a multivariable prediction model for individual prognosis or diagnosis (tripod): the tripod statement.
\newblock {\em Annals of Internal Medicine}, 162(1):55--63, 2015.

\bibitem{Austin2020}
P.~C. Austin, F.~E. Harrell~Jr, and D.~van Klaveren.
\newblock Graphical calibration curves and the integrated calibration index (ici) for survival models.
\newblock {\em Statistics in Medicine}, 39(21):2714--2742, 2020.

\bibitem{Nagrani2021}
A.~Nagrani, S.~Yang, A.~Arnab, A.~Jansen, C.~Schmid, and C.~Sun.
\newblock Attention bottlenecks for multimodal fusion.
\newblock In {\em Advances in Neural Information Processing Systems}, volume~34, pages 14200--14213, 2021.

\bibitem{Singh2022}
L.~Singh, R.~R. Janghel, and S.~P. Sahu.
\newblock A hybrid feature fusion strategy for early fusion and majority voting for late fusion towards melanocytic skin lesion detection.
\newblock {\em International Journal of Imaging Systems and Technology}, 32(4):1231--1250, 2022.

\bibitem{Mauricio2023}
J.~Maurício, I.~Domingues, and J.~Bernardino.
\newblock Comparing vision transformers and convolutional neural networks for image classification: A literature review.
\newblock {\em Applied Sciences}, 13(9):5521, 2023.

\bibitem{Howard2019}
A.~Howard, M.~Sandler, G.~Chu, L.~C. Chen, B.~Chen, M.~Tan, and H.~Adam.
\newblock Searching for mobilenetv3.
\newblock In {\em Proceedings of the IEEE/CVF international conference on computer vision}, pages 1314--1324, 2019.

\bibitem{He2016}
K.~He, X.~Zhang, S.~Ren, and J.~Sun.
\newblock Deep residual learning for image recognition.
\newblock In {\em Proceedings of the IEEE conference on computer vision and pattern recognition}, pages 770--778, 2016.

\bibitem{lu2023mi-zero}
Ming~Y Lu, Bowen Chen, et~al.
\newblock Visual language pretrained multiple instance zero-shot transfer for histopathology images.
\newblock In {\em CVPR}, 2023.

\bibitem{zhou2024keep}
Xiao Zhou, Luoyi Sun, et~al.
\newblock A knowledge-enhanced pathology vision-language foundation model for cancer diagnosis.
\newblock {\em arXiv preprint arXiv:2412.13126}, 2024.

\bibitem{chen2022pancancer}
Richard~J Chen, Ming~Y Lu, et~al.
\newblock Pan-cancer integrative histology-genomic analysis via multimodal deep learning.
\newblock {\em Cancer Cell}, 40(8):865--878, 2022.

\bibitem{lu2021aipathology}
Ming~Y. Lu, Tiffany~Y. Chen, Drew F.~K. Williamson, Ming~Y. Zhao, Mahmoud Shady, Jana Lipkova, and Faisal Mahmood.
\newblock Ai-based pathology predicts origins for cancers of unknown primary.
\newblock {\em Nature}, 594(7861):106--110, 2021.

\bibitem{bejnordi2017camelyon16}
Babak~Ehteshami Bejnordi, Mitko Veta, Paul Johannes~van Diest, Bram van Ginneken, Nico Karssemeijer, Geert Litjens, et~al.
\newblock Diagnostic assessment of deep learning algorithms for detection of lymph node metastases in women with breast cancer.
\newblock In {\em JAMA}, volume 318, pages 2199--2210. American Medical Association, 2017.

\bibitem{bandi2018detection}
Peter Bandi, Oscar Geessink, Qaiser~F Manson, Marije~C Van~Dijk, Markus Balkenhol, Meyke Hermsen, et~al.
\newblock Detection of lymph node metastases in breast cancer from whole-slide histopathology images using deep learning: participation in the camelyon17 challenge.
\newblock In {\em Medical Image Analysis}, volume~56, pages 30--43. Elsevier, 2019.

\bibitem{kather2019predicting}
Jakob~Nikolas Kather, Jan Krisam, Piroon Charoentong, and et~al.
\newblock Predicting survival from colorectal cancer histology slides using deep learning: A retrospective multicenter study.
\newblock {\em PLoS Medicine}, 16(1):e1002730, 2019.

\end{thebibliography}

\end{document}